\documentclass[prd,showpacs,amsmath,showkeys,twocolumn,floatfix]{revtex4} 
\usepackage{epsfig,dcolumn}
\usepackage{graphicx}
\usepackage[usenames]{color}
\usepackage{graphicx}
\usepackage{bm}
 \usepackage{ifpdf}

\def\lsim{\mathrel{\rlap{\lower4pt\hbox{\hskip1pt$\sim$}}
    \raise1pt\hbox{$<$}}}
\def\gsim{\mathrel{\rlap{\lower4pt\hbox{\hskip1pt$\sim$}}
    \raise1pt\hbox{$>$}}}
\begin{document}

\title{Amplitudes for the analysis of the decay $J/\psi \to K^{+} K^{-} \pi^0$ }

\author{  Peng Guo$^1$,  Ryan Mitchell$^1$, Matthew Shepherd$^{1}$  and Adam P. Szczepaniak$^{1,2}$}
\affiliation{ $^1$ Physics Department, Indiana University, Bloomington, IN 47405, USA. \\
$^2$Center For Exploration  of Energy and Matter, Indiana University, Bloomington, IN 47408, USA. 
}

\date{\today}

\begin{abstract} 
We construct an analytical  model for two channel, two-body scattering amplitudes,  and then apply it in the description of the three-body  $J/\psi \to K^+K^-\pi^0$ decay. In the construction of the partial wave amplitudes,  we combine the low energy resonance region with the Regge asymptotic behavior determined from direct two-body production.   We find that resonance production in the $K\pi$ channel in $J/\psi$ decays seems to differ from that observed in direct $K\pi$ production, while the mass distribution in the $K\bar K$ channel may be compatible. 
    \end{abstract}

\maketitle

\section{Introduction}
\label{intro} 

Meson spectroscopy has played an important role in developing  phenomenology and  gaining insights  into QCD in the non-perturbative domain.  In an amplitude analysis of experimental data it is necessary to explore all of the available theoretical constraints because the extraction of resonance  parameters  requires the analysis of partial waves outside of the kinematic range of experimental data. In particular, amplitudes describing the mass distribution of a two-body subsystem 
 in a quarkonium decay may be different from those describing scattering of the same two particles. 
   In this paper we focus on the  isospin $1/2,1$ and spin-one,  $P$-wave scattering amplitudes in the $\pi\pi$, $K{\bar K}$ and $K\pi$ channels and compare the phase shift data with  two-body mass distributions from the  three-body $J/\psi \to K^+K^-\pi^0$ decay.  These amplitudes are dominated by the ground state vector resonances $\rho(770)$ and $K^*(892)$ that are well established as quark-antiquark, QCD bound states that are weakly coupled to the meson-meson continuum. There is also strong experimental evidence for higher mass vector resonances,  although precisely how many and to what extent these  are related to QCD single hadron states remains an open issue~\cite{Diekmann:1988,Hyams:1973,Aston:1980,Bisello:1989,Donnachie:1987}.

In Table \ref{latticemass} we list the masses of the 
  lowest vector meson states obtained from recent lattice QCD simulations  \cite{Jozef:2010} and the 
    quark potential model  \cite{Isgur:1985} and compare them to the data compiled by 
     the   Particle Data Group~(PDG)~\cite{PDG}.   Below $2\mbox{ GeV}$ the PDG lists two excited isovector resonances, the $\rho'(1450)$ and $\rho''(1700)$, that could  have the quark model assignments of $2S$ and $1D$, respectively.   In the lattice simulations of  \cite{Jozef:2010} the   average pion mass is approximately $400\mbox{ MeV}$, which puts the $\rho$ meson approximately 
    $130\mbox{ MeV}$ above its measured mass.  Shifting the vector mesons masses 
     from lattice computations  down by $130\mbox{ MeV}$ puts 
      the  first excited state  around $1600$~MeV, which is $\sim 150\mbox{ MeV}$ higher than the measured mass of the $\rho'(1450)$.  While a resonance in the $1600-1700\mbox{ MeV}$ mass range can be clearly inferred from the $\pi\pi$ scattering phase shift data ~\cite{Hyams:1973}, the experimental  evidence for the $\rho'(1450)$ is  ambiguous~\cite{PDG}.   The main motivation for the $\rho'(1450)$ comes from the need to accommodate data on $4\pi$ 
       production~\cite{Achasov:1997,Achasov:2002}. To the best of our knowledge, however,  there has been no comprehensive analysis of all available $P$-wave data and the importance 
        of the various inelastic channels, possibly even the dominant one, 
         $K\bar K$~\cite{Hyams:1973} is yet to be settled. 
    For example, an alternative scenario that seems to be supported by the lattice results  
       might be that the $2S$ and $1D$ states are above $1.6\mbox{ GeV}$ while 
         any residual strength corresponding  to  the PDG $\rho'(1450)$ could due to residual interactions between pions  and/or inelastic channel effects.

   \begin{center}
\begin{table}[htdp]
\begin{center}
\begin{tabular}{|c|c|c|}
\hline
&$\rho(1^{--})$ & $K^{*}(1^{-})$ \\
\hline
                               &0.90       &  0.95 \\
Lattice QCD  \cite{Jozef:2010} &1.8        &  1.8 \\
                               &$\cdots$   &  $\cdots $ \\
\hline
                               & 0.77($1^{3}S_{1}$)  & 0.90($1^{3}S_{1}$)  \\ 
Quark Model \cite{Isgur:1985}  & 1.45($2^{3}S_{1}$)  & 1.58($2^{3}S_{1}$)  \\ 
                               & 1.66($1^{3}D_{1}$)  & 1.78($1^{3}D_{1}$)  \\ 
                               & $\cdots$            & $ \cdots$ \\
\hline 
                               & 0.775  & 0.895 \\
 PDG \cite{PDG} 	       & 1.465  & 1.414 \\
                               & 1.720  & 1.717 \\
  \hline
\end{tabular}
\end{center}
\caption{Masses of the first few lowest-lying  vector meson resonances. \label{latticemass}}
\label{default}
\end{table}
\end{center}

\begin{figure}
\begin{center}
\includegraphics[width=3.0 in,angle=0]{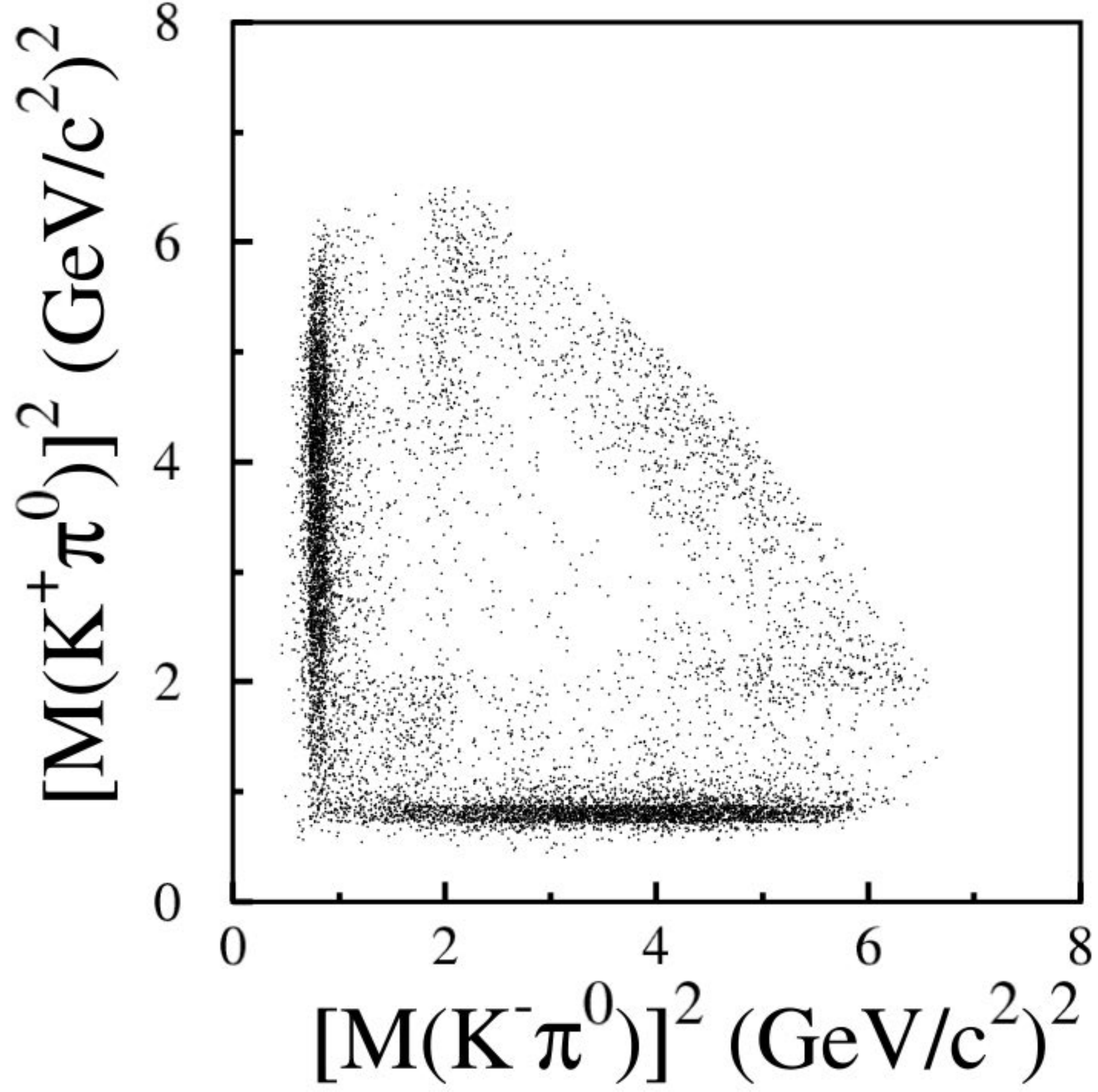}  
\caption{The $J/\psi \to K^+K^- \pi^0$  Dalitz plot  distribution  from the BESII Collaboration~\cite{BES:2006}.
\label{dbes2}}
\end{center}
\end{figure}

The vector mesons discussed above can be produced in  $J/\psi \to K\bar K \pi$ and $3\pi$ decays. In this paper we  focus on the former; we studied the latter in Ref.~\cite{Guo:2010}. 
 The $J/\psi \to K\bar K \pi$ decay has been analyzed by the BESII Collaboration~\cite{BES:2006}.  
The Dalitz plot distribution of the  $K^{+} K^{-} \pi^0$ events  has  clearly visible sharp bands corresponding to the isospin-$1/2$,  $K^{\pm *}(892)$ and weaker bands in the first excited $K'^*$ resonance region. The distribution is shown in Fig.~\ref{dbes2}. 
There is  also  a significant enhancement  in the low $K^{+} K^{-}$ invariant mass region.  In the BESII analysis this broad band was associated with a new isovector  $P$-wave   resonance, the $X(1570)$,  decaying to $K^+K^-$ with the pole position at  $(1576- 409 i)\mbox{ MeV}$ seen through a strong destructive  interference with the  $\rho(1700)$. 
  There have been several theoretical attempts to explain this result~\cite{Liu:2007,Li:2007}. 

In this work we address the following questions.  Can the broad enhancement in the low-mass $K{\bar K}$ channel be described by the $P$-wave $K{\bar K}$ amplitudes determined  from phase shift analysis?  And, more generally, can the Dalitz plot distribution of $K{\bar K}\pi$ events in the $J/\psi$ decay be described in terms of $K{\bar K}$ and $K\pi$ amplitudes reconstructed from phase shift analysis? To do so, we use,  and further develop 
   (by incorporating asymptotic energy dependence), 
        the $P$-wave $\pi\pi$ and $K{\bar K}$ amplitudes initially constructed in \cite{Szczepaniak:2010re,Guo:2010}.  The amplitudes that we use  have the correct analytical properties, 
     satisfy two body unitarity and reproduce the known data on $\pi\pi$ 
     scattering~\cite{Hyams:1973,Protopopescu:1973,Estabrooks:1974}. 
  In  \cite{Guo:2010}  we successfully used these amplitudes  to describe the $J/\psi \to \pi^+\pi^-\pi^0$ Dalitz distribution. In particular we have found that, since the $\rho''(1700)$ is quite  inelastic~\cite{Hyams:1973}, destructive interference with the virtual $J/\psi \to K{\bar K} \pi \to 3\pi$ process is important in reducing the Dalitz plot intensity in the  $\rho''$ resonance region. We will investigate if it is possible that a similar phenomena is in operation in the $K{\bar K} \pi$ final state and whether   the virtual $J/\psi \to 3\pi \to K{\bar K}\pi$ decay may be responsible 
    for the broad structure at low $K^{+}K^{-}$ invariant mass.

 This paper is organized as follows.    The partial wave decomposition of the decay $J/\psi \to K^+K^- \pi^0$ is given in Sec.~\ref{sec:P}. 
In   Sec. \ref{kkanalyticalmodel}, we discuss our $P$-wave amplitudes and compare with the BESII data of Fig.~\ref{dbes2}.   Ideally, the set of partial waves that are developed here could be used in a full Dalitz plot analysis, but this requires a full knowledge of experimental acceptances and resolutions. In this work we simply compare, qualitatively, a sample of Dalitz plot distributions generated from our amplitudes with the BESII result of Fig.~\ref{dbes2}.  We include more details on the amplitude construction in the appendices.

\section{ Partial wave amplitudes in the  $J/\psi \rightarrow  K^{+} K^{-} \pi^{0}$ decay } 
\label{sec:P} 

Denoting the four momenta by $p_{\pm,0}$, $P$ for $K^\pm$, $\pi^0$  and $J/\psi$ respectively, the general expression for the $J/\psi\to K^{+} K^{-} \pi^0$ amplitude is given by, 
\begin{equation} 
\langle \pi^0 K^+K^-,out|J/\psi(\lambda),in\rangle =  i (2\pi)^4  \delta^4\left(\sum_{i=0,\pm} p_i - P\right) 
 T_\lambda.   \label{J}
\end{equation} 
 The Dalitz  plot invariants are defined by $s_{ij} = (p_i + p_j)^2$ with $i,j = \pm,0$ referring to $K^\pm$ and the $\pi^0$, respectively. The general expression for the helicity amplitude of $T_\lambda$ is given by 
 \begin{widetext} 
\begin{eqnarray} 
\label{t} 
T_\lambda =  \sum_{S,L} \sum_{\mu=\pm,0} N_{SL\mu}  [ D^{1*}_{\lambda, \mu}(r_{+-}) d^S_{\mu,0}(\theta^{+}_{+-})   F^{+-}_{SL}(s_{+-}) 
+ D^{1*}_{\lambda, \mu}(r_{+0}) d^S_{\mu,0}(\theta^{+}_{+0})   F^{+0}_{SL}(s_{+0}) + D^{1*}_{\lambda, \mu}(r_{-0}) d^S_{\mu,0}(\theta^{-}_{-0})   F^{-0}_{SL}(s_{-0}) ]
\end{eqnarray}  
\end{widetext}
where $N_{SL\mu} = \sqrt{3(2S+1)}  \langle S \mu; L0| 1\mu \rangle/4\pi$.   Here $\lambda$ is the spin projection of the $J/\psi$ along the $e^+e^-$ beam axis, which together with $x$ and $y$ define a lab coordinate system, $S$ is the spin of a two particle subsystem (the isobar),  and $L$ is the relative orbital angular momentum between the isobar and the spectator meson.     The  rotation  $r_{ij}$ is given by three Euler angles, $r_{ij}=r_{ij}(\phi_{ij},\vartheta_{ij},\psi_{ij}^{i})$   which rotate the standard configuration  in  the $(ij)k$ coupling scheme, 
   to the actual one.   In  the  standard configuration of the $(ij)k$ coupling $J/\psi$ is at rest,  particle  $k$  has momentum along the negative $z$ axis, and particles $i$ and $j$ have momenta in the $xz$ plane with the particle $j$ moving in the positive  $x$  direction. 
 The azimuthal and polar angles,   $\phi_{ij}$ and $\vartheta_{ij}$,  are defined in the $J/\psi$ rest 
   frame and refer to the actual direction of motion of the   $(ij)$  pair. 
 Finally, $\psi_{ij}^{i}$ and $\theta_{ij}^{i}$ are the azimuthal and the polar angle of the $i$-th particle in the $(ij)$, two-particle (isobar) rest frame. 

The scalar form factors $F^{ij}_{S L}(s_{ij})$ describe the dynamics of the decay in the isobar model 
    {\it i.e.} under the assumption that in a given isobar channel  the form factors are  functions of the sub-energy of that isobar only.    In the $L-S$ basis, the    parity of the $K^{+} K^{-} \pi^{0}$ state is given by $P=
   (-1)^{S+L+1}$ and  under charge conjugation the two isobar channels,    
   $|(K^{+}\pi^{0})K^{-} \rangle$ and  $|(K^{-}\pi^{0})K^{+}\rangle$  are exchanged  while the third isobar channel,  $|(K^{+}K^{-})\pi^{0} \rangle$ is a charge conjugation eigenstate with the eigenvalue 
   $(-1)^{S}$. Thus charge conjugation invariance implies that  in  Eq.~(\ref{t})   there  are only two independent  form factors which we define as, 
  \begin{eqnarray}
F^{+-}_{SL}\equiv \frac{1-(-1)^{S}}{2}F^{K \bar{K}}_{SL}, \ \ F^{+0}_{SL} = - F^{-0}_{SL} \equiv -F^{K\pi}_{SL}
\end{eqnarray}
and  obtain, 
  \begin{widetext} 
\begin{eqnarray} 
T_\lambda &=&  \sum_{S,L} \sum_{\mu=\pm 1,0}  N'_{SL\mu}   [ D^{1*}_{\lambda, \mu}(r_{+-}) d^S_{\mu,0}(\theta^{+}_{+-})  F^{+-}_{SL}(s_{+-}) -D^{1*}_{\lambda, \mu}(r_{+0}) d^S_{\mu,0}(\theta^{+}_{+0})    F^{K\pi}_{SL}(s_{+0}) + D^{1*}_{\lambda, \mu}(r_{-0}) d^S_{\mu,0}(\theta^{-}_{-0})    F^{K\pi}_{SL}(s_{-0}) ] \nonumber \\
\end{eqnarray}  
\end{widetext}
with  $N'_{SL\mu} \equiv N_{SL\mu} (1+(-1)^{S+L})/2$. The $\mu=0$ component vanishes  due to  parity conservation  and we can further reduce the partial wave expansion to 
  \begin{widetext} 
\begin{eqnarray} 
T_\lambda &=&  \sum_{S,L} N'_{SL1} \{ [ D^{1*}_{\lambda, 1}(r_{+-})+D^{1*}_{\lambda, -1}(r_{+-}) ] d^S_{1,0}(\theta^{+}_{+-})  
 \frac{1-(-1)^{S}}{2}
  F^{K \bar{K}}_{SL}(s_{+-}) \nonumber  \\
&-&[ D^{1*}_{\lambda, 1}(r_{+0}) +D^{1*}_{\lambda, -1}(r_{+0})  ]d^S_{1,0}(\theta^{+}_{+0})    F^{K\pi}_{SL}(s_{+0}) +[ D^{1*}_{\lambda, 1}(r_{-0}) +D^{1*}_{\lambda, -1}(r_{-0})  ]d^S_{1,0}(\theta^{-}_{-0})    F^{K\pi}_{SL}(s_{-0}) \}.
\end{eqnarray}  
\end{widetext}
Finally, it is useful to rewrite the above amplitude in terms of a single set of angles describing orientation of the decay plane.  Using the relation between Euler rotations, 
   \begin{equation} 
   r_{+-} = r_{-0} r(0,\chi_+,0) = r_{+0} r^{-1}(0,0,\pi)r^{-1}(0,\chi_-,0),
   \end{equation} 
   where $\chi_+ (\chi_-)$ is the angle between $K^+$ ($K^-$)  and $\pi^0$ and in the $K^{+} K^{-}\pi^{0}$ rest  frame enables to write $T$ in terms of $r_{+-}$ alone
    \begin{widetext} 
\begin{eqnarray} \label{helicityamp}
T_\lambda = \sum_{S,L} N'_{SL1} [ D^{1*}_{\lambda, 1}(r_{+-})+D^{1*}_{\lambda, -1}(r_{+-})] 
   [ d^S_{1,0}(\theta^{+}_{+-}) \frac{1 - (-1)^S}{2} F^{K \bar{K}}_{SL}(s_{+-}) +d^S_{1,0}(\theta^{+}_{+0})   F^{K\pi}_{SL}(s_{+0})+ d^S_{1,0}(\theta^{-}_{-0})   F^{K\pi}_{SL}(s_{-0}) ]. \nonumber \\
\end{eqnarray}  
\end{widetext}
 The  allowed quantum numbers in the $K^{+}K^{-}$ channel are $S^{PC}=1^{- -} (\rho),3^{--}(\rho_{3}), \cdots$, and  in  the $K^{\pm}\pi^{0}$ channels,  $S^{P} = 1^{-}(K^{*}), 2^{+}(K^{*}_{2}), 3^{-} (K^{*}_{3}),\cdots$. In the following we will assume that the Dalitz distribution can be saturated with the lowest  partial waves, {\it i.e.} $P$-wave in both $K^{+}K^{-}$ and $K^{\pm}\pi^{0}$ channels,  
 and we test this hypothesis by studying the effect of the $D$-wave resonances in the $K\pi$ channels.  Parity  conservation implies $S=L$;  therefore, in the following we will simply denote   $F^{ij}_{S L}$  by $F_{L}^{ij}$. The (unnormalized) $J/\psi$ partial decay width with respect to one of the Dalitz  invariants  ({\it e.g.} $M_{K^+K^-} =\sqrt{s_{+-}}$)   is obtained by integrating the square of the decay amplitude over the orientation of the decay plane and the other independent invariant,  
 \begin{equation}
\frac{d \Gamma}{d\sqrt{ s_{+-}}} = N  \sqrt{ s_{+-} }
  \int^{s^{up}_{-0}(s_{+-})}_{s^{dn}_{-0}(s_{+-})}  d s_{-0} |T |^{2} , \label{pwa} 
 \end{equation}
 and
 \begin{eqnarray}
&  &|T|^2=   |\sum_{S,L} N'_{SL1}  [ d^S_{1,0}(\theta^{+}_{+-})   \frac{1-(-1)^{S}}{2}F^{K \bar{K}}_{L}(s_{+-})  \nonumber \\
&+& d^S_{1,0}(\theta^{+}_{+0})   F^{K\pi}_{L}(s_{+0})+ d^S_{1,0}(\theta^{-}_{-0})   F^{K\pi}_{L}(s_{-0}) ]|^2. \label{pwaamplitude} 
\end{eqnarray}  
The overall normalization ($N$) is adjusted to match the measured number of events. 
It is $|T|^2$ that 
 determines the distribution of events in the  Dalitz  plot, ({\it i.e.} $|T|^2 = const.$ would give    
   a flat distribution).
The integration limits,    $s^{up/dn}_{-0}(s_{+-})$ are roots of the equation which define the boundary of the Dalitz plot, 
\begin{widetext}
\begin{eqnarray}
 s_{+-} s_{+0} s_{-0} - ( s_{+0} + s_{-0})(m_{\pi}^{2}m_{K}^{2} + M^{2} m_{K}^{2}) -s_{+-}(m_{K}^{4}+ M^{2} m_{\pi}^{2}) + 2(m_{K}^{4}    m_{\pi}^{2} + M^{2} m_{K}^{4}   + 2 M^{2} m_{K}^{2} m_{\pi}^{2} ) =0. 
\end{eqnarray}
\end{widetext}
 Projections along $M_{K^+\pi} = \sqrt{s_{+0}}$ and $M_{K^-\pi} = \sqrt{s_{-0}}$ axis   can be defined analogously.   
 
  In the following we discuss parameterizations of the form factors $F_{L}^{K\bar K}$ and $F_{L}^{K\pi}$ in terms of two-body amplitudes. Any  parameters remaining in these parameterizations, which are related to the production process as opposed to final state interactions should be determined by fitting the Dalitz distributions. As discussed in Sec.~\ref{intro} we do not fit the published Dalitz distribution, but instead show the predicted  distributions for specific values of these parameters.


\section{ Theoretical  Model for Form Factors} 
  \label{kkanalyticalmodel}

  Unitarity relates production form factors to two-body amplitudes. 
    In ~\cite{Szczepaniak:2010re, Guo:2010} we constructed analytical representations for the  isovector, $P$- wave, two-body, $\pi\pi$ and $K{\bar K}$ amplitudes. Here we further extend the analysis of ~\cite{Guo:2010}  by constraining the high energy behavior, and extend the approach to the $K\pi$ channel.       
        We begin with a $K$-matrix, phenomenological  parameterization of  the known data (on the real axis) on phase shifts and elasticities. Even though the $K$-matrix offers an analytical representation for the amplitude, it often leads to spurious poles and zeros of the amplitude when extrapolated outside the physical region. Therefore we use the analytical representation 
   for phase shifts and inelasticity via  the $K$-matrix  only in the data region and smoothly 
    extrapolate  to match with the asymptotic behavior of the partial waves at high energies. 
  We then use the amplitudes constructed this way over the whole  physical energy range  
   as input into the Omn\'es-Muskhelishvili integral to construct the part of the scattering 
    amplitude  regular on the left side of the complex $s$-plane. With the $N(s)/D(s)$ representation, which is described below, we determine the amplitude over the entire $s$-plane. Finally we solve 
   the unitarity relation for the form factors and write 
 the $J/\psi$  decay amplitude  in terms of the denominator functions $D(s)$ and production functions $c_\alpha(s)$.  In the following we describe these steps in a little more detail.  
 All details of the amplitude construction are given in the Appendix. 

\subsection{Amplitude Parameterization} 

       In ~\cite{Guo:2010} to describe the high energy 
   limit of the isovector $P$-wave,  the following hypothesis was made: the $S$ matrix is saturated by two channels, $\pi\pi$ and $K{\bar K}$ and the elastic channel phase shifts asymptotically  approach   a multiple of $\pi$ with elasticity $\eta$ approaching $1$. Even though $J/\psi$ decays probe only a limited energy range, and are quite insensitive to details of the asymptotic behavior we might as well use a different  hypothesis  that is better rooted in high energy phenomenology.
It is known that at high energies, elastic cross sections slowly grow with energy  almost approaching the Froissart  bound.  This implies that at impact parameter larger than the interaction region $O(1\mbox{ fm})$ there is  
 no interaction while the low partial waves are suppressed as if scattering from a "gray disk." 
  The low partial waves correspond to  $L  <<   L_0(s)$ where $L_0(s) \sim  \sqrt{s}/2  \mbox{ fm}$  
   and while the interaction radius grows logarithmically with energy, the scattering of the low partial waves becomes logarithmically suppressed, {\it i.e.} $\eta_L \sim 1 - O(1/\log s)$. 
  In the language of Regge exchanges this picture corresponds to the Pomeron exchange at high energies. Furthermore, since asymptotically the number of inelastic channels grows rapidly, each individual  inelastic amplitude, 
  {\it e.g} $\pi\pi \to K{\bar K}$ is expected to fall off with energy,  and is represented by 
  exchange of non-vacuum quantum numbers, aka meson Regge trajectories. 
  The hypothesis of two-channel dominance in the high energy limit is therefore not necessarily 
   well justified    and in the following we adopt the Regge picture of high energy scattering. 
      Matching the $K$-matrix  parameterization  of the low energy data with Regge asymptotics, leads to amplitudes  of the form (we drop the angular momentum label on the partial wave),   
   
   \begin{eqnarray}
   \label{phases} 
  &&t_{\alpha,\beta}(s) = |t_{\alpha,\beta}(s)| e^{i\phi_{\alpha,\beta}(s)}
    =  \left\{ \begin{array}{c} t_{\alpha,\beta}^{Kmatrix}(s)  , s < s_{low} \nonumber    \\ 
t_{\alpha\beta}^{Regge}(s) ,  s> s_{high}   \end{array}\right.  \nonumber \\
   \end{eqnarray}
 with
 $ t_{\alpha,\beta}^{Kmatrix}(s)$ and $t_{\alpha\beta}^{Regge}(s)$ 
  determined from  $K$-matrix fits to the low energy data and Regge fits to the high energy fixed $t$-data, respectively.  Greek indices denote two body channels, {\it i.e.} $\alpha = (i,j) = \pi\pi,K\bar K$, {\it etc.} 
 For energies between $s_{low}$ and $s_{high}$, we  smoothly connect both real and imaginary parts     of the $K$-matrix and Regge amplitudes. 
    The denominator function in the $N/D$ parameterization 
   
   \begin{equation} 
   t_{\alpha\beta}(s) = \frac{N_{\alpha\beta}(s)}{D_{\alpha\beta}(s)} \label{noverd} 
   \end{equation} 
is then obtained  from the phase of the scattering amplitude using the Omn\'es-Muskhelishvili  
solution of the unitarity relation ($s_{th}  \ge min(s_\alpha,s_\beta)$ where $s_\alpha$ is the  $\alpha$-channel threshold) 
\begin{equation} 
\frac{ \mbox{Im} D_{\alpha\beta}(s)}{D_{\alpha\beta}(s)} = - \sin\phi_{\alpha\beta}(s) e^{-2i\phi_{\alpha\beta}(s)}
\end{equation} 
 and is  given by 
   \begin{equation} 
      D_{\alpha\beta}(s) = e^{- \frac{s}{\pi} \int_{s_{th}} ds' \frac{\phi_{\alpha\beta}(s')}{s' (s' - s)} } \label{dgen} 
   \end{equation} 
   where we conveniently normalized $D_{\alpha\beta}(0)=1$. 
    The numerator functions $N_{\alpha\beta}(s)$ are given by the largely unknown discontinuity of the amplitudes on the left hand cut.  For the purpose of solving the unitarity relation for the $J/\psi$ decay form factors,  which will be discussed below ({\it cf.}  Eq.~(\ref{fu})), it is convenient  to have $N_{\alpha\beta}(s)$'s for all intervening $\alpha,\beta$ channels  having the same analytical form. This is certainly a simplifying approximation, nevertheless we have found that with a simple parameterization 
    \begin{equation} 
    N_{\alpha\beta}(s) = \frac{\lambda_{\alpha\beta}}{s + s_L}  \label{nc} 
    \end{equation} 
 and  with the two-body amplitudes given by Eqs.~(\ref{noverd}),(\ref{dgen}), it is indeed possible to obtain good fits to the two-body scattering data, {\it i.e.} phase shifts and elasticity. 
 
 Having constructed the two-body amplitudes, the next step is to relate them to the production form factors. This is done through the unitarity relations which relate the 
 imaginary part of the form factors to the two-body amplitudes, 
        \begin{equation} 
  \label{fu} 
  \mbox{Im} \hat F^\alpha_L(s) =  \sum_\beta  t^*_{\alpha,\beta}(s) \rho^\beta (s) \hat F^\beta_{L}(s)  
   \end{equation} 
with $_{\alpha\beta}t$ representing the elastic $L$-partial wave scattering amplitude between two-body channels 
 $\alpha=(ij)$ and $\beta = (i'j')$. $\hat F$  is the reduced form factor (with the barrier factor removed), 
   \begin{equation} 
    F_L^{ij}(s) = q^{L}_{ij}(s) p^{L}_{k}(s)  \hat F^{ij}_{L}(s), 
    \end{equation} 
 with $q_{ij}(s)$  being the relative momentum between mesons $i$ and $j$, 
 \begin{equation} 
\label{qij} 
  q_{ij}(s) =\sqrt{ \frac{[s-(m_{i}+m_{j})^{2}][s-(m_{i}-m_{j})^{2}]}{4 s}} 
  \end{equation} 
   and 
      \begin{equation} \label{splitmomentum}
        p_{k}(s) =  \sqrt{ \frac{[s-(M+m_{k})^{2}][s-(M-m_{k})^{2}]}{4 M^{2} }},  
  \end{equation}
 with $M$ being the $J/\psi$ mass,  the relative momentum between the $(ij)$ pair and the spectator  meson $k$. $\rho_{\alpha}(s) = 2q_{ij}/\sqrt{s}$ describes the two-particle phase space. 
 It is straightforward to show  that if the scattering amplitude is dominated by a single resonance
 below inelastic threshold ($\rho = \rho^\alpha$, $\rho^\beta = 0$ for $\beta \ne \alpha$)  the solution of the unitarity  condition for $\hat F$ is
\begin{equation} 
\hat F^{\alpha}_L(s) = c(s) BW^{L}_R(s)  = \frac{c(s)}{m_{R}^{2} - s - i m_{R} \Gamma_L (s) }, 
\end{equation} 
where $BW^L_R(s)$ is the Breit-Wigner amplitude (with an energy dependent width 
$\Gamma_L(s)$)
 and  $c(s)$ is a real polynomial in $s$.  In  the general multiple-channel case, with the two-body  amplitudes all parameterized by the same numerator function, as in Eq.~(\ref{nc}) 
the solution to Eq.~(\ref{fu}) is given by~\cite{Pham:1976yi},   
 \begin{equation} 
 \hat  F^\alpha(s) = \sum_\beta \frac{c_\beta(s)}{D^{\alpha\beta}(s)}  \label{fdall}
 \end{equation} 
 with $c_\beta(s)$ being analytic functions in the right hand plane and $\mbox{Im} c_\alpha(s) = 0$ for $s > 0$. 
 
 \subsection{Results} 
 
\begin{figure}[hh]
\begin{center}
\includegraphics[width=3 in,angle=0]{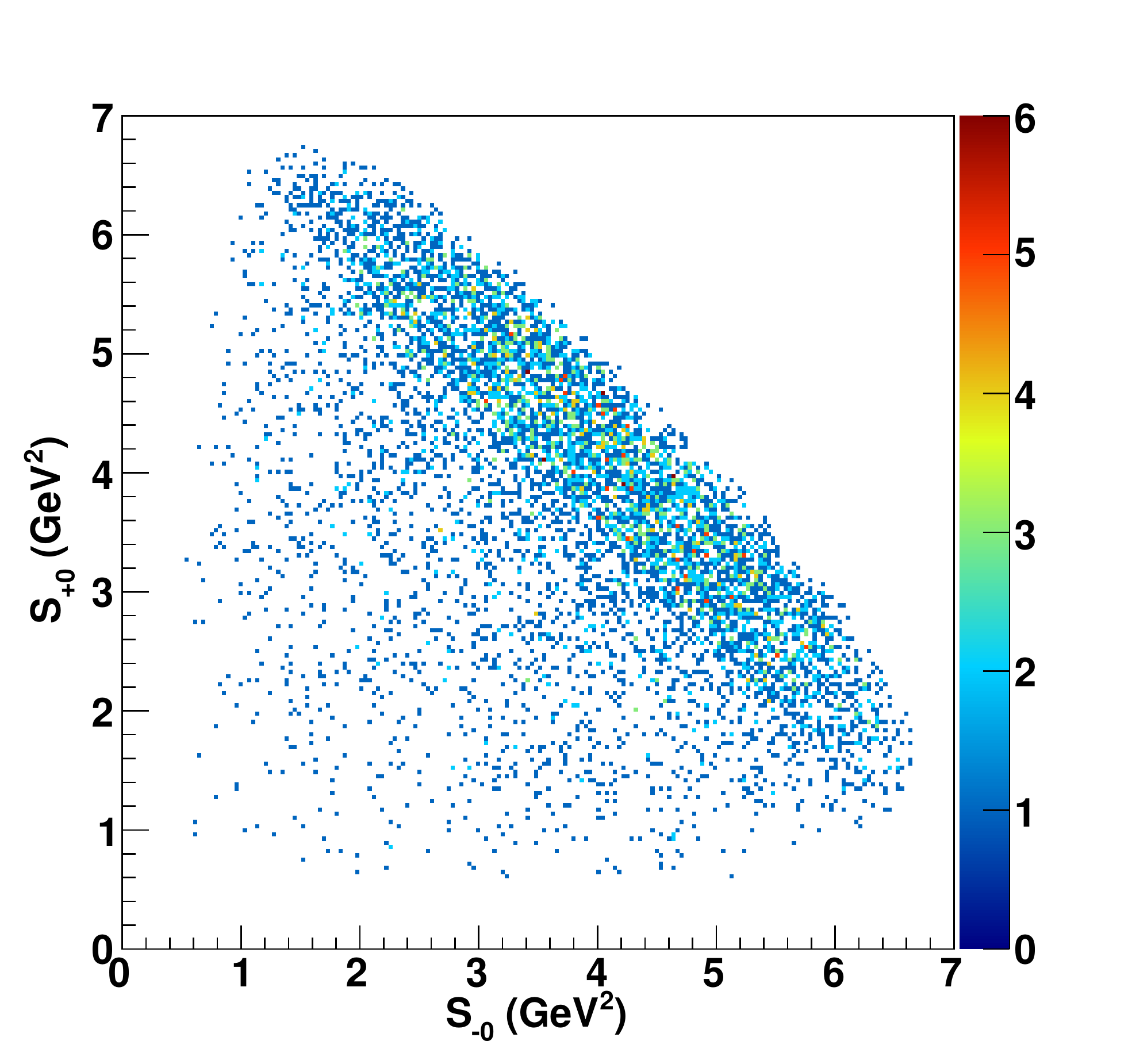}  
\includegraphics[width=3 in,angle=0]{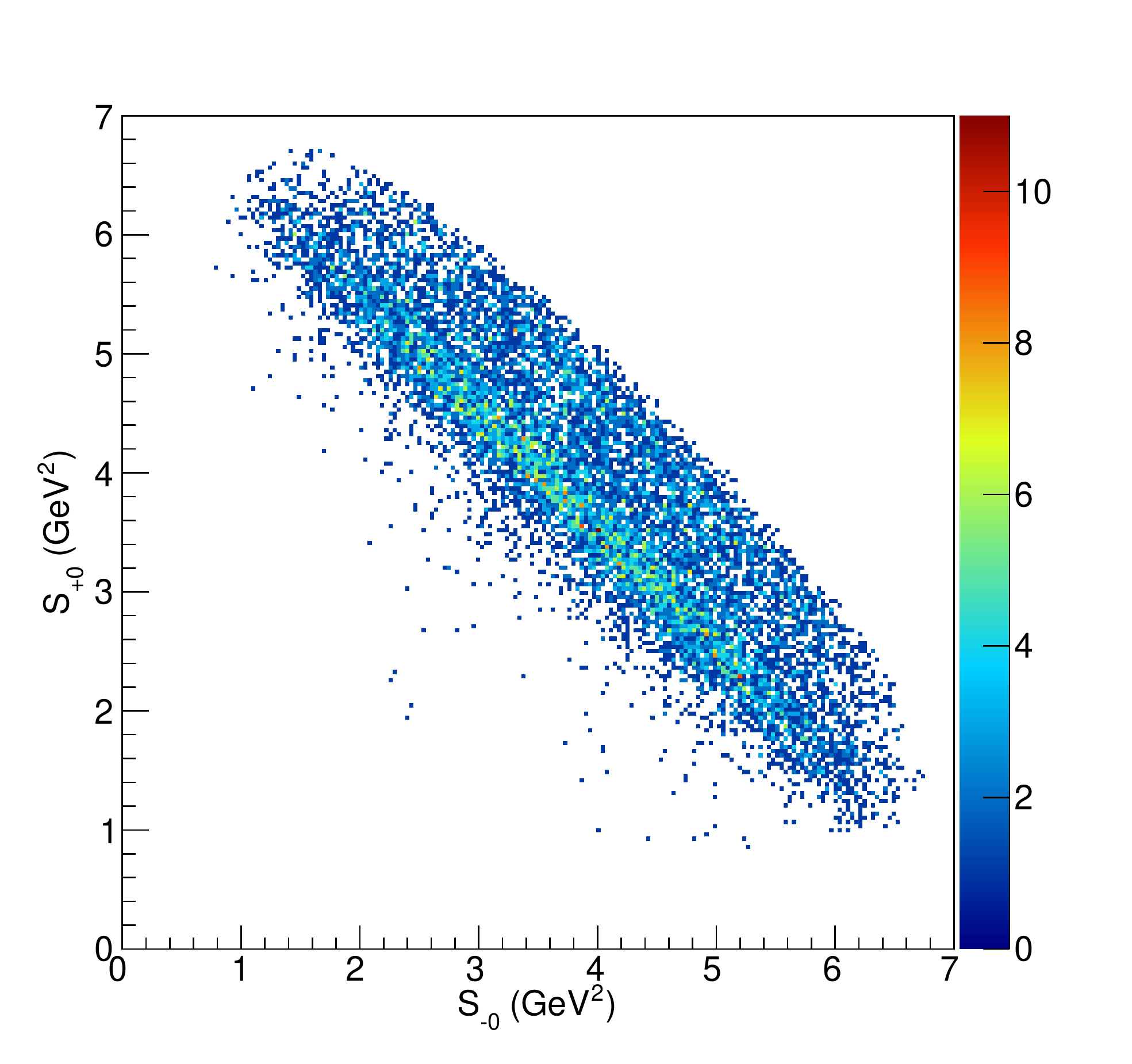}  
\caption{ Dalitz plot distribution obtained using, in Eq.~(\ref{fdall}), a single two-body 
$K\bar K\to K\bar K$ amplitude (top) and a single two-body  $\pi\pi \to K\bar K$ amplitude (bottom) ({\it i.e.}  with $c_{K\bar K}=1$ ($c_{\pi\pi}=1$) for the top (bottom)  and $c_\alpha(s) = 0$ for all other waves).    
\label{dkkpp}}
\end{center}
\end{figure}

 As discussed in Sec.~\ref{intro} the original BESII analysis was based on the isobar, resonance parameterization of all three two-body channels. In the absence of a known isovector $P$-wave $K{\bar K}$ resonance to describe the low mass $K{\bar K}$ enhancement, it was necessary to introduce a new resonance, the $X(1570)$. The $\pi\pi$ phase isovector $P$-wave shift data, however, points to significant inelasticity above $1.6\mbox{ GeV}$, which following \cite{Hyams:1973} we have attributed to the $K{\bar K}$ channel. The effect of the coupled  $\pi\pi$ and $K{\bar K}$  channels on the  $K^+K^- \pi^0$ mass distribution which follows from Eq.~(\ref{fdall})   is shown in Fig.~\ref{dkkpp}. 

\begin{figure}[hh]
\begin{center}
\includegraphics[width=3 in,angle=0]{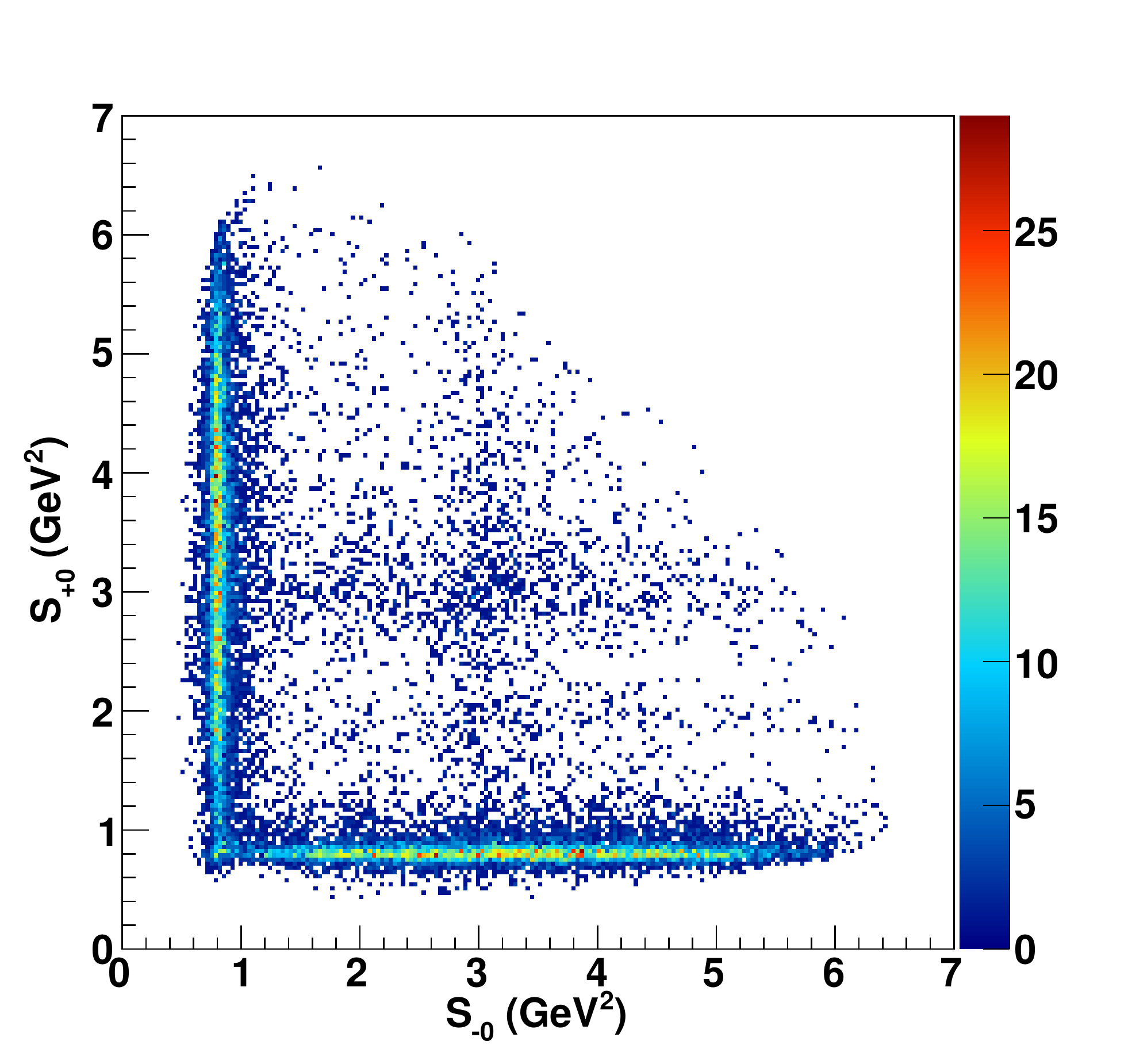}  
\caption{ As in Fig.~\ref{dkkpp} but with a single two-body, $K\pi \to K\pi$ 
 amplitude. \label{dkp}}
\end{center}
\end{figure}

In Fig.~\ref{dkp} we show the Dalitz distribution obtained using the single $K\pi$ channel amplitude 
 (details discussed in the Appendix).  Besides the $K^*(892)$ peaks, bands at $M_{K\pi} = 1.75 \mbox{ GeV}$  are  clearly visible in both $K^+ \pi^0 $ and $K^-\pi^0$   mass projections.  These are 
  due to the $K^*(1680)$ resonance clearly  seen in the $K\pi$ phase shift 
  analysis~\cite{Aston:1988,Mercer:1971,Estabrooks:1978}
  but apparently not so in the $K\pi$ production from the 
 $J/\psi$ decay ({\it cf.} Fig.~\ref{dbes2}). This clear discrepancy indicates that 
  it is  not sufficient to use a single channel $K\pi$ amplitude in the parameterization of the corresponding    form factor in the $J/\psi$ decay. As discussed in the Appendix the $K\pi$ amplitude is indeed inelastic above  $M_{K\pi} \sim 1.5 \mbox{ GeV}$ with a possibility of a large coupling to the $K^*(892)\pi$ channel. 

\begin{figure}[hh]
\begin{center}
\includegraphics[width=3 in,angle=0]{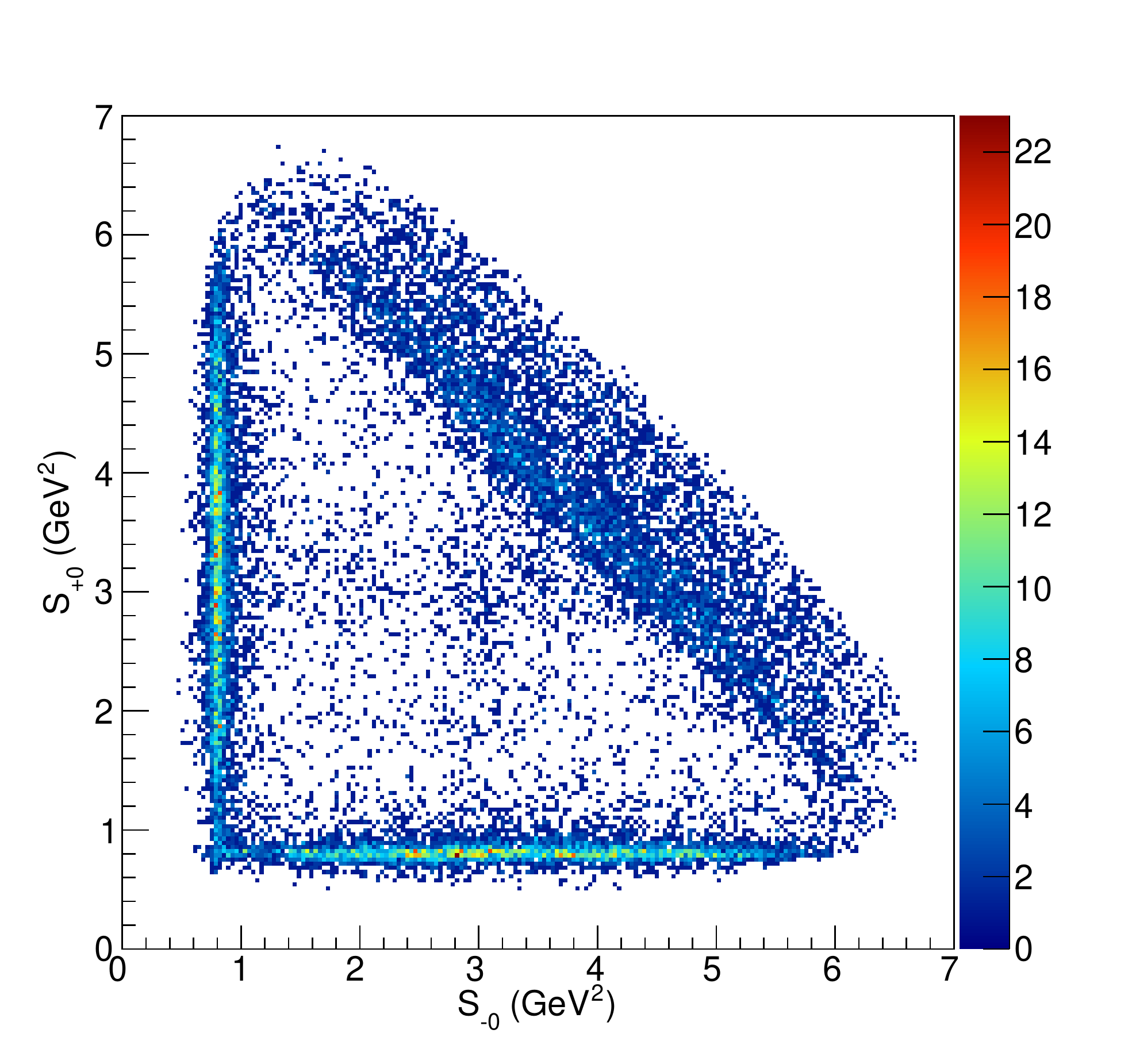}  
\caption{ As in Fig.~\ref{dkkpp} but with three amplitudes, 
 $K{\bar K} \to K{\bar K}$, $\pi\pi \to K{\bar K}$ and $K\pi \to K\pi$ with relative production 
coefficients satisfying $c_{K\pi}(s):c_{K\bar K}(s):c_{\pi\pi}(s) = 1: 0.3:-0.7$. 
 \label{all}}
\end{center}
\end{figure}

Finally, in Fig.~\ref{all} we show the Dalitz distribution obtained with a combination of three amplitudes, $K{\bar K} \to K{\bar K}$, $\pi\pi \to K{\bar K}$ and $K\pi \to K\pi$ with relative production 
coefficients, $c_\alpha(s)$ chosen to best match the observed distribution in Fig.~(\ref{dbes2}). While the low mass $K\bar K$ region seems to be fairly  well described, the resonance structures in he $K\pi$ channel do not match between the elastic $t_{K\pi \to K\pi}$ and $J/\psi$ decay amplitude.  
  The  $\pi \pi \to K\bar{K}$ and $K\bar{K} \to K\bar{K}$  amplitudes behave rather smoothly in the 
   region corresponding to the  $K\pi$ resonances 
    and  do not give enough strength to reducing the peak  from the second $K^*$ resonance region. 
    Thus we anticipate that the discrepancy is due to inelasticities in the $K\pi$ channel itself. 
  Since we are only comparing Dalitz distributions as opposed to fitting data, 
 we do not attempt to further improve the comparison. 
    It is worth noting that the $K^{*}(1410)$  listed in the PDG 
  is  indeed quite inelastic with only a $6.6\%$ branching to $K \pi$. 

\section{Discussion and conclusion} \label{discussion}
Based on unitarity and analyticity we have constructed  a set of analytical two-body amplitudes, which implement the known phase shift data. 
These extend our previous work 
  in coupled channel $P$-wave $\pi \pi$ and $K\bar{K}$ systems and the $J/\psi \to 3\pi$ 
  decay~\cite{Guo:2010}. 
The two-body amplitudes are only an approximation to the three-body decay, nevertheless they provide a useful starting point and should match below inelastic thresholds. 
       We compared the analysis of the $J/\psi$ decay  with these amplitudes to the original analysis 
       of the BESII collaboration,  which was based on the isobar model with coherent Breit-Wigner resonances.   The isobar model with the known, low mass 
        resonances only and without inelasticities 
     cannot faithfully produce  the broad structure of low $K^{+}K^{-}$ invariant mass, which is why 
       in the BESII analysis an additional $P$-wave resonance  $X(1576)$ coupled to $K^+K^-$ was introduced. Our preliminary study indicates that the $K\bar K$ low-mass region may be described by the inelasticity in the $\pi\pi \to \pi\pi$ wave if attributed to the coupling between $\pi\pi$ and $K\bar K$ channels.  A single $K\pi \to K\pi$ amplitude is strongly affected by the second vector $K^*(1680)$  resonance as observed in $K\pi$ phase shift analysis. However, in $J/\psi$ decay this resonance seems to be suppressed. It is worth noting that a similar suppression of the first excited 
       isovector-vector resonance is also observed in the $3\pi$ decay of $J/\psi$~\cite{Guo:2010}.

\section{Acknowledgments}
The authors would like to thank Mikhail Gorchtein for helpful discussion. 
This work was  supported in part by the  US Department of Energy grant under
contract DE-FG0287ER40365, National Science Foundation PIF grant number 0653405.

\appendix

\section{Analytical model for the $P$-wave  isovector $\pi\pi \to \pi\pi$, $\pi\pi\to K \bar{K}$  and $K{\bar K} \to K{\bar K}$ amplitudes} \label{appendkmat}
 
 \subsection{$K$-matrix parameterization, ($s < s_{low}$)} 
  \label{kmatorg}
 
    \begin{figure}
\begin{center}
\includegraphics[width=3 in,angle=00]{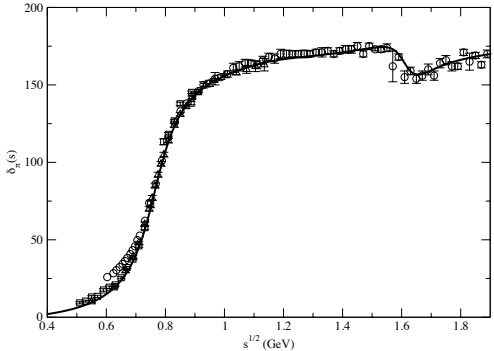}  
\caption{Phase shift of the $P$-wave $\pi \pi$ amplitude. Data is taken from 
 \cite{Hyams:1973} (circles) ,\cite{Protopopescu:1973} (triangles) , and \cite{Estabrooks:1974} (squares). 
   The solid line  is the result of the fit  to $\delta_\pi$ and $\eta$  with the analytical $K$-matrix representation described  in the text. 
\label{fig:phase}}
\end{center}
\end{figure}

    \begin{figure}
\begin{center}
\includegraphics[width=3.0 in,angle=00]{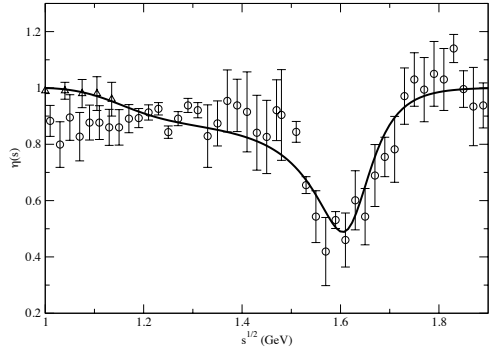}  
\caption{Same as in Fig.~\ref{fig:phase} for the inelasticity $\eta$ \label{fig:eta}}
\end{center}
\end{figure}

We use a two channel $K$-matrix~\cite{Guo:2010} to fit the data  on $\pi\pi \to \pi\pi$ $P$-wave phase shift and elasticity $\eta$ from~\cite{Hyams:1973,Protopopescu:1973,Estabrooks:1974} (Fig.~\ref{fig:phase},\ref{fig:eta}). With the $S$-matrix saturated by two channels, the model makes a prediction 
 for the phase shift in the $K{\bar K} \to K{\bar K}$ channel.  In this section $\alpha,\beta = \pi, K$
  correspond to the two body channels $\pi\pi$ and $K{\bar K}$, respectively. 
 The  2-channel $K$-matrix representation is given by
\begin{equation} 
[ \hat t^{-1}(s)]_{\alpha\beta} = [K^{-1}(s)]_{\alpha\beta} + \delta_{\alpha\beta}  (s - s_\alpha)I_\alpha(s),   \label{tk}
\end{equation} 
where 
\begin{equation} 
I_\alpha(s) = I_\alpha(0) - \frac{s}{\pi} \int_{s_\alpha}^\infty ds'   \sqrt{1 - \frac{s_\alpha}{s'}} \frac{1}{ (s' - s)s'}.  
\end{equation} 
A convenient choice for the subtraction constant, $I_\alpha(0)$, is to take $\mbox{Re}I_\alpha(M^2_\rho) = 0$ so that one of the poles of  $K_{\pi\pi}$  corresponds to the Breit-Wigner mass squared, $M^2_\rho=(0.77\mbox{ GeV})^2$,  of  the $\rho$ meson. Using the general two-pole parameterization of the $K$ matrix, 
 \begin{eqnarray}
& & K_{\pi\pi} =  \frac{ \alpha_\pi^2 }{M_{\rho}^{2}-s}   +\frac{\beta_{\pi}^{2} }{s_{2}-s} + \gamma_{\pi \pi}, \; 
K_{KK} =  \frac{ \beta^{2}_{K}}{s_{2}-s} + \gamma_{KK} \nonumber \\ 
& & K_{\pi K}  = K_{K\pi} =  \frac{\beta_{\pi} \beta_{K}}{s_{2}-s} + \gamma_{\pi K} ,
 \end{eqnarray}
 where $\alpha_\pi^2 = \Gamma_{\rho}M^2_{\rho}/(M_\rho^2 - s_\pi)^{3/2}$ and fitting 
  the $P$-wave $\pi\pi$ phase shift, $\delta_\pi$, and the elasticity, $\eta$, we obtain 
  $\Gamma_{\rho} = 0.140 \mbox{ GeV}$, and 
\begin{eqnarray} \label{tkparam}
&&   \sqrt{ s_{2}}  = 1.4708 \mbox{ GeV},  \ \ \ \  \beta_{\pi} = 0.199, \ \ \ \  \beta_{K} =0.899,
 \nonumber  \\  && \gamma_{\pi \pi} = 5.62\times 10^{-2}, \ \ \ \ \ \gamma_{\pi K} = 0.104, \ \ \ \ \ \gamma_{KK} = 1.525, \nonumber \\
 \end{eqnarray}
  with the $\gamma$'s in units of $\mbox{ GeV}^{-2}$. The comparison of the phase shift and the inelasticity obtained with this parameterization with the data is shown in Fig.~\ref{fig:phase},\ref{fig:eta}.
 To illustrate  unphysical features of the $K$-matrix parameterization we rewrite Eq.~(\ref{tk})  
   using the standard $N/D$ representation for  $\hat{t}_{\alpha\beta} = t_{\alpha\beta}/( 4 q_{\alpha} q_{\beta} )$.  With the normalization  $D_{\alpha\beta}(0)=1$ we obtain, 
 \begin{widetext} 
 \begin{eqnarray} 
&& N_{\pi\pi}(s) = \lambda_{\pi\pi} 
\frac{s - z_{\pi\pi}}{(s - s_{L,1})(s - s_{L,2})}, \; D_{\pi\pi}(s) = \exp\left(-\frac{s}{\pi} 
\int_{s_\pi}ds' \frac{\phi_{\pi\pi}(s')}{s' (s' - s)}\right), \nonumber \\
& & N_{\pi K}(s) =
\frac{ \lambda_{\pi K} }{(s - s_{L,1})(s - s_{L,2})}, \; D_{\pi K}(s) = \frac{ s_{1,\pi K} s_{2,\pi K} }{ (s -s_{1,\pi K} )(s - s_{2,\pi K})}
\exp\left(-\frac{s}{\pi} \int_{s_\pi}ds' \frac{\phi_{\pi K}(s')}{s' (s' - s)}\right), \nonumber \\
& & N_{KK}(s) =  \lambda_{KK} 
\frac{s - z_{KK}}{(s - s_{L,1})(s - s_{L,2})}, \; D_{KK}(s) = \exp\left(-\frac{s}{\pi} \int_{s_\pi}ds' 
\frac{\phi_{KK}(s')}{s' (s' - s)}\right),   \label{KND}
\end{eqnarray}
\end{widetext}
with $\lambda_{\pi\pi} = 5.649$, $\lambda_{KK} = 2.271 $ and $\lambda_{\pi K} = 3.048\mbox{ GeV}^2$.  In this $K$-matrix model, the left hand cut of $N$ is reduced to two poles at $s_{L,1} = -13.87\mbox{GeV}^2$ and $s_{L,2} = -0.787\mbox{ GeV}^2$, respectively. There are also  first order   zeros  in $N_{\alpha\beta}$ at 
 $z_{\pi\pi} = -0.867\mbox{ GeV}^2$ and  $z_{KK} = -13.78\mbox{GeV}^2$.  
  Above the $K{\bar K}$ threshold  the phase of the inelastic amplitude $\phi_{\pi K}$ is given by $\phi_{\pi K} = \delta_\pi + \delta_K$. From the $K$ matrix we find that,  asymptotically, $\phi_{\pi K}(\infty) = 2\pi$, which corresponds to two CDD poles: one at the $\rho$ mass, $s_{1,\pi K} = M_\rho^2$, and the other at   $s_{2,\pi K} =  s_2 + \beta_\pi\beta_K/\gamma_{\pi K} = 3.884\mbox{ GeV}^2$.  Thus, while the $K$~matrix parameterization faithfully reproduces the $\pi\pi$ phase shift and elasticity in the whole available energy range, from $\pi\pi$ threshold  up to $1.9\mbox{ GeV}$,  extrapolation  beyond this range is problematic.  The rapid decrease of $\phi_{\pi\pi}$ around $s \sim 6\mbox{ GeV}^2$ seems unphysical.
 In  the $\pi\pi \to K{\bar K}$ channel, the two CDD poles 
  at $m_\rho^2$ and $s_2 + \beta_\pi\beta_K/\gamma_{\pi K}$ are clearly an artifact  of the 
   pole  parameterization of the $K$-matrix.  A CDD pole in the inelastic channel above threshold  ({\it cf.} the pole at $s_{2,\pi K} = 3.884\mbox{ GeV}^2$)  leads to a discontinuity  in a phase shift and is unphysical. It also implies vanishing inelasticity, $\eta=1$ at this energy.    A pole between $\pi\pi$ and $K{\bar K}$ thresholds is admissible, {\it e.g.} the pole at $s_{1,\pi K} = m_\rho^2$, but its strict overlap with the $\rho$ mass is also an  artifact of the parameterization.  Since the  phase space available in $J/\psi$ decay extends  up to $s_{\pi\pi} \sim 9 \mbox{GeV}^2$ we need to remove these unphysical features of the $K$-matrix amplitude. As discussed in 
   Sec.~\ref{kkanalyticalmodel} we do this by using the $K$-matrix amplitudes below $s_{low}$ and above $s_{high}$ we will use Regge parameterization.

\subsection{Regge parameterization ($s > s_{high}$) }   \label{appendregge}
\begin{figure}[hh]
\begin{center}
\includegraphics[width=3 in,angle=0]{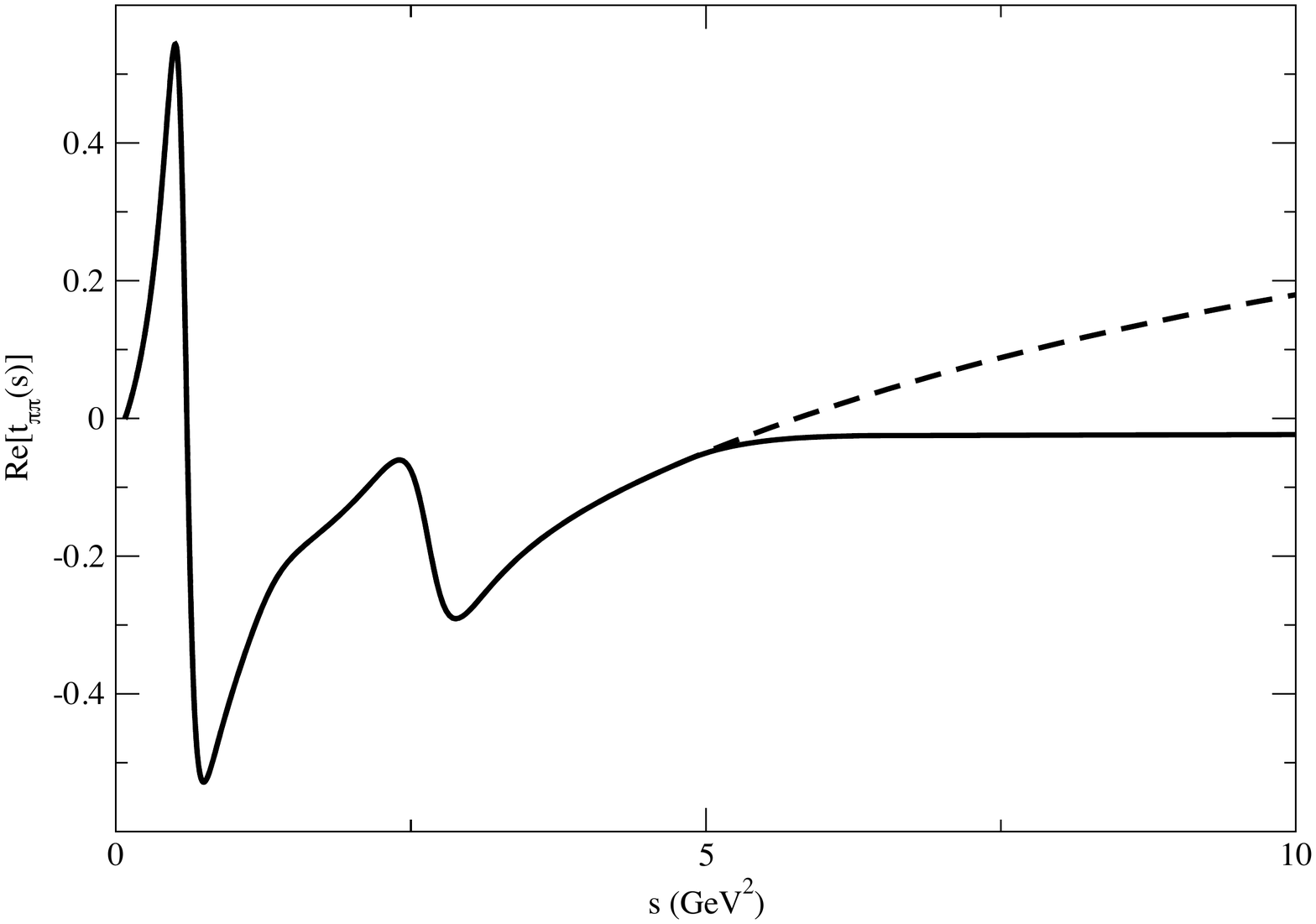} 
\includegraphics[width=3 in,angle=0]{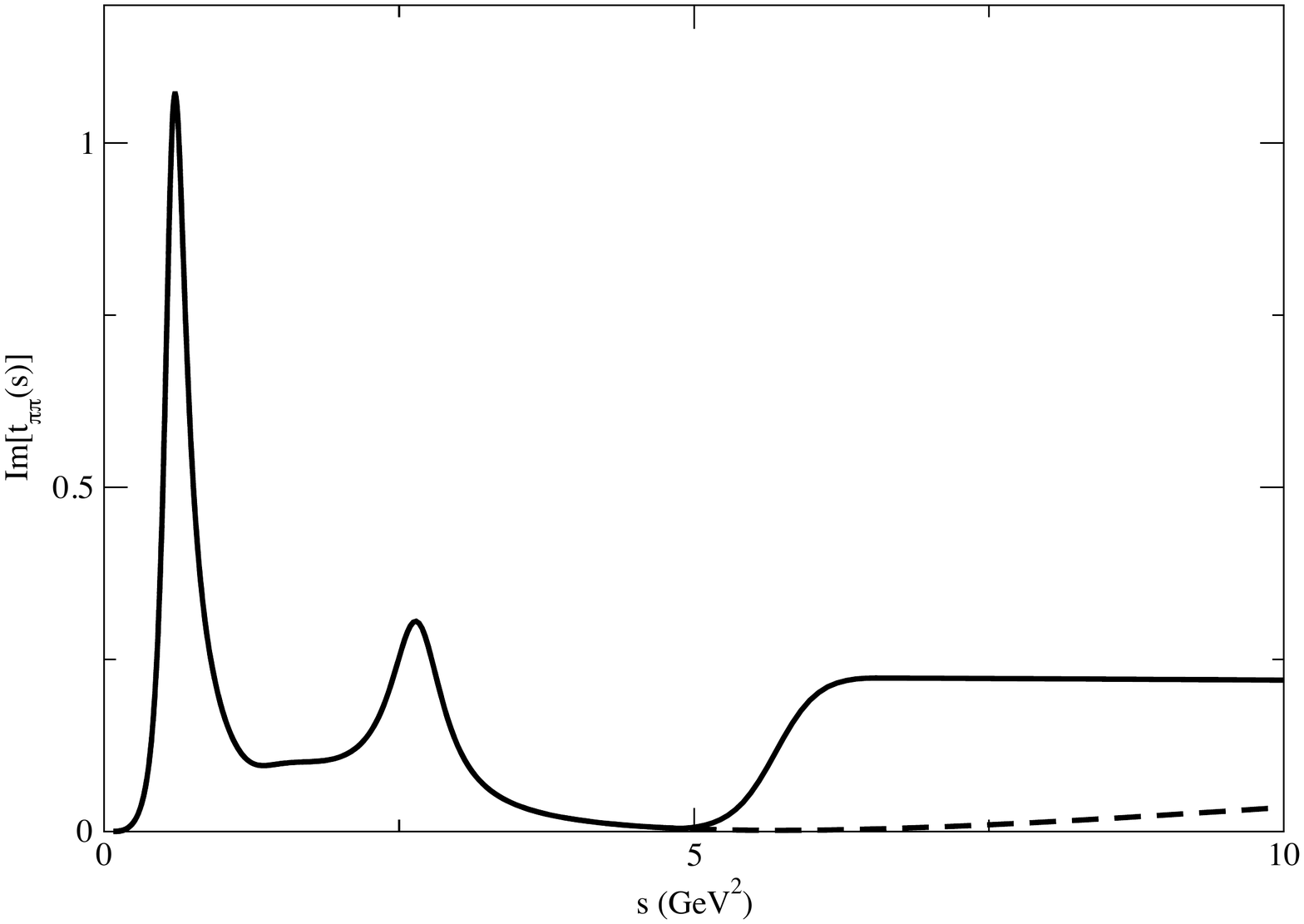}  
\caption{  Real (top) and imaginary (bottom) parts of the isovector, $P$-weave amplitude, 
 $t_{\pi \pi}(s)$ (solid lines). Dashed line is the result of the $K$-matrix parameterization. 
\label{tpipiregge}}
\end{center}
\end{figure}

 Regge analysis of   $\pi \pi \rightarrow \pi \pi$ scattering has been studied  recently 
  in \cite{pelaez:2004,Ananthanarayan:2001,pelaez:2005} and here  we use the results of ~\cite{pelaez:2005}.  Parameters in Regge amplitudes were constrained by analyzing $NN$, $\pi N$ and $\pi \pi$ scattering data.  For completeness we give the relevant formulas below. 
      \begin{itemize} 
\item{} $\pi\pi \to \pi\pi$ 
 \end{itemize} 
 Regge parameterization involves the Regge poles 
     corresponding to $t$-channel exchange of  the  Pomeron($P$), the $P'$ (associated with the  $f_{2}(1270)$ trajectory) and  the $\rho$. The $t$-channel isospin amplitudes are given by 
 \begin{eqnarray} 
&& F^{(I_{t}=0)}_{\pi \pi}(t,s,u)  \nonumber \\
& = &   - \frac{1+e^{-i \pi \alpha_{P}(t)}}{\sin \pi \alpha_{P}(t)}P(s,t) - \frac{1+e^{-i \pi \alpha_{P'}(t)}}{\sin \pi \alpha_{P'}(t)} P'(s,t), \label{pomeron} \nonumber \\
&& \\
 && F^{(I_{t}=1)}_{\pi \pi}(t,s,u)  \nonumber \\
   & =  &  \frac{1-e^{-i \pi \alpha_{\rho}(t)}}{\sin \pi \alpha_{\rho}(t)}\beta_{\rho}  \frac{1+ \alpha_{\rho}(t)}{1+ \alpha_{\rho}(0)} [1+ d_{\rho} t ]e^{bt}   (s/\hat{s})^{\alpha_{\rho}(t)}, \label{f2regge}  \nonumber \\
   && \\
 &&\mbox{Im} F^{(I_{t}=2)}_{\pi \pi}(t,s,u)   =  \beta_{2}  e^{bt}  (s/\hat{s})^{\alpha_{\rho}(t)  + \alpha_{\rho}(0)-1}  \label{rhoregge}.  \nonumber \\
 \end{eqnarray}
 where ($\hat s =1\mbox{ GeV}$) 
 \begin{eqnarray}
 P(s,t) &=& \beta_{P} \alpha_{P}(t) \frac{1+\alpha_{P}(t)}{2} e^{bt}(s/\hat{s})^{\alpha_{P}(t)}, \label{pomeronparam}  \\
 P'(s,t) &=& \beta_{P'}   \frac{ \alpha_{P'}(t)[1+\alpha_{P'}(t)]}{\alpha_{P'}(0)[1+\alpha_{P'}(0)]} e^{bt}(s/\hat{s})^{\alpha_{P'}(t)} ,   \ \ \  \ \ 
  \end{eqnarray}
and the trajectories are   given by 
 \begin{eqnarray} \label{pomerontraject}
  &&\alpha_{P}(t)=\alpha_{P}(0)+ t \alpha'_{P},  \nonumber \\
   &&\alpha_{P'}(t) =  \alpha_{\rho}(t)= \alpha_{\rho}(0) + t \alpha'_{\rho} +\frac{1}{2} t^{2} \alpha''_{\rho} .  
  \end{eqnarray}
   Numerical values of all parameters are given in Eqs.~(B5), (B6)  of  \cite{pelaez:2005}. The $s$-channel isospin, partial wave amplitudes are normalized according to 
   \begin{eqnarray} \label{norm}
F^{(I_{s})}_{\alpha,\beta}(s,t,u)=(\sqrt{2})^{\sigma} \frac{4}{\pi} \sum_{L}(2L+1) t^{(LI_s)}_{\alpha,\beta}(s)  P_{L}(\cos \theta),  \nonumber \\
\end{eqnarray}
where $(\sqrt{2})^{\sigma}$ is the identical particle symmetry factor:  $\sigma =2$ for   
  $\pi \pi \leftrightarrow \pi \pi$, $\sigma=1$   for $\pi \pi \leftrightarrow K \bar{K}$ and  $\sigma=0$
    for $ K \bar{K}\leftrightarrow  K \bar{K}$.  The $s$-channel amplitudes with  $I_s=0,2$ are symmetric under $t \leftrightarrow u$ exchange, and the $I_s=1$ amplitude is antisymmetric and 
       $s \leftrightarrow t$ crossing leads to the following relation between the $s$ and  the $t$-channel isospin amplitudes,

 \begin{eqnarray}
  F^{(I_{s}=1)}_{\pi \pi}(s,t,u) &=& \frac{1}{3}  F^{(I_{t}=0)}_{\pi \pi}(t,s,u)  + \frac{1}{2}  F^{(I_{t}=1)}_{\pi \pi}(t,s,u)  \nonumber \\
&-& \frac{5}{6}  F^{(I_{t}=2)}_{\pi \pi}(t,s,u) - (t\rightarrow u) .  
 \end{eqnarray}
The $(t\leftrightarrow u) $ exchange brings in the u-channel Regge poles
 (these were ignored in \cite{pelaez:2005} where only the forward $t = 0$ limit was considered). 
  Finally, projecting out the $P$-wave amplitude yields, 
 \begin{eqnarray} \label{tregge}
&& t^{Regge}_{\pi \pi}(s)= \frac{\pi}{16}\int_{-1}^{1} (d \cos  \theta)   \cos \theta  \nonumber \\
&\times&   [\frac{1}{3}  F^{(I_{t}=0)}_{\pi \pi}(t,s,u)  + \frac{1}{2}  F^{(I_{t}=1)}_{\pi \pi}(t,s,u) \nonumber \\
& -& \frac{5}{6}  F^{(I_{t}=2)}_{\pi \pi}(t,s,u) - (t\rightarrow u) ]. 
\end{eqnarray}
The angular integration is done numerically. The leading asymptotic behavior due to Pomeron exchange can be calculated analytically and is given by, 
\begin{eqnarray}
&& t^{Regge}_{\pi \pi}(s) \simeq i \frac{\pi}{16}  \frac{1}{3}  \int_{-1}^{1}(d \cos  \theta)   \cos \theta   [  P(s,t)  -  P(s,u) ]  \nonumber \\
& \simeq& i   \frac{\pi}{24} \beta_{P} \frac{- 3 \alpha'_{P} +2 (b+\alpha'_{P} \ln s)  }{(b+\alpha'_{P} \ln s)^{2}}  s^{\alpha_{P}(0)-1} .   
\end{eqnarray}
To combine the $K$-matrix ($s < s_{low}$) with the Regge projected ($s>s_{high}$) amplitudes 
 into  the full $P$-wave $\pi\pi \to \pi\pi$ amplitude, 
 \begin{eqnarray}
  &&t_{\pi \pi}(s)  =  \left\{ \begin{array}{c} t_{\pi \pi}^{Kmatrix}(s)  , s < s_{low}    \\ 
t_{\pi \pi}^{Regge}(s) ,  s> s_{high}   \end{array}\right., 
   \end{eqnarray}
   we choose $\sqrt{s_{low}} = 2.20 \mbox{ GeV}$ and $\sqrt{s_{high}} = 2.56 \mbox{ GeV}$, 
       and use a simple analytical formula to  smoothly join the two amplitudes between 
        $s_{low}$ and $s_{high}$.  The result is shown in Fig.~\ref{tpipiregge}.

\begin{itemize} 
\item{} $\pi\pi \to K{\bar K}$ 
\end{itemize} 
Asymptotically the $t$-channel  amplitude it is dominated by the $K^*$ trajectory, 
\begin{eqnarray} \label{Kstarregge}
&&F^{(I_{t} = \frac{1}{2})}_{\pi K }(t,s,u)  \nonumber \\
&=&   \frac{1- e^{ - i \pi \alpha_{K^{*}}(t)}}{ \sin  \pi  \alpha_{K^{*}}(t) } \beta_{K^{*}} \frac{2 \alpha_{K^{*}}(t)+1}{2 \alpha_{K^{*}}(0)+1} e^{bt} (\alpha'_{K^{*}} s)^{\alpha_{K^{*}}(t)}. \nonumber  \\
\end{eqnarray}
Following  \cite{Ananthanarayan:2002} we use 
 $b=2.4$ $GeV^{-2}$, and  $\alpha_{K^{*}}(t) =0.352+\alpha'_{K^{*}}  t $ with 
$\alpha'_{K^{*}}  = 0.882GeV^{-2}$. The $s$-channel $I_s= 1/2$ amplitude is antisymmetric 
 under $t \leftrightarrow u$ exchange, 
\begin{eqnarray}
 F^{(I_{s}=1)}_{ \pi K}(s,t,u) = - F^{(I_{s}=1)}_{ \pi K}(s,u,t) . 
 \end{eqnarray}
 and from $s \leftrightarrow t$ crossing we obtain, 
\begin{eqnarray}
 && F^{(I_{s}=1)}_{ \pi K}(s,t,u) \nonumber  \\
 &=& \frac{2}{3}F^{(I_{t} = \frac{1}{2})}_{\pi K }(t,s,u)  - \frac{2}{3}F^{(I_{t} = \frac{3}{2})}_{\pi K}(t,s,u)   - (t\rightarrow u). \nonumber \\
 \end{eqnarray}
In terms of  $F^{(I_{s}=1)}_{ \pi K}(s,t,u)$  the properly normalized $P$-wave in $\pi\pi \to K{\bar K}$ is finally given by 
 \begin{eqnarray} \label{pipitokknorm}
&& t^{Regge}_{\pi K}(s) =\frac{\pi}{8  \sqrt{2}} \int_{-1}^{1} (d \cos \theta )\cos \theta  \nonumber \\
&\times& \left[  \frac{2}{3}F^{(I_{t} = \frac{1}{2})}_{\pi K }(t,s,u) -(t \rightarrow u)\right]. 
 \end{eqnarray}
 We can fix $\beta_{K^{*}}$ by matching our formula in Eq.~(\ref{Kstarregge}) to Eq.~(81) in \cite{Ananthanarayan:2002} in the limit of $t\rightarrow 0$ (forward direction). Taking into 
 account differences in  normalization  employed here and  used in  \cite{Ananthanarayan:2002}, we find 
\begin{eqnarray}
&& \mbox{Im} F^{(I_{t} = \frac{1}{2})}_{\pi K } (t,s,u) |_{s \rightarrow \infty, t \rightarrow 0}    \nonumber \\
&=&   \beta_{K^{*}}  (\alpha'_{K^{*}} s)^{\alpha_{K^{*}}(0)}  =  \frac{3}{4 \pi  }  \frac{ \lambda}{\Gamma[\alpha_{K^{*}} (0)]} (\alpha'_{K^{*}}s)^{\alpha_{K^{*}}(0)} \nonumber \\
\end{eqnarray}
with $\lambda =1.82$   taken from  \cite{Ananthanarayan:2002}   and 
\begin{eqnarray}
  \beta_{K^{*}}=  \frac{3}{4 \pi  }   \frac{ \lambda}{\Gamma[\alpha_{K^{*}} (0)]}      =  0.172 .
\end{eqnarray}
 Asymptotically, $ t^{Regge}_{\pi K}(s)$ approaches 
 
\begin{figure}[hh]
\begin{center}
\includegraphics[width=3 in,angle=0]{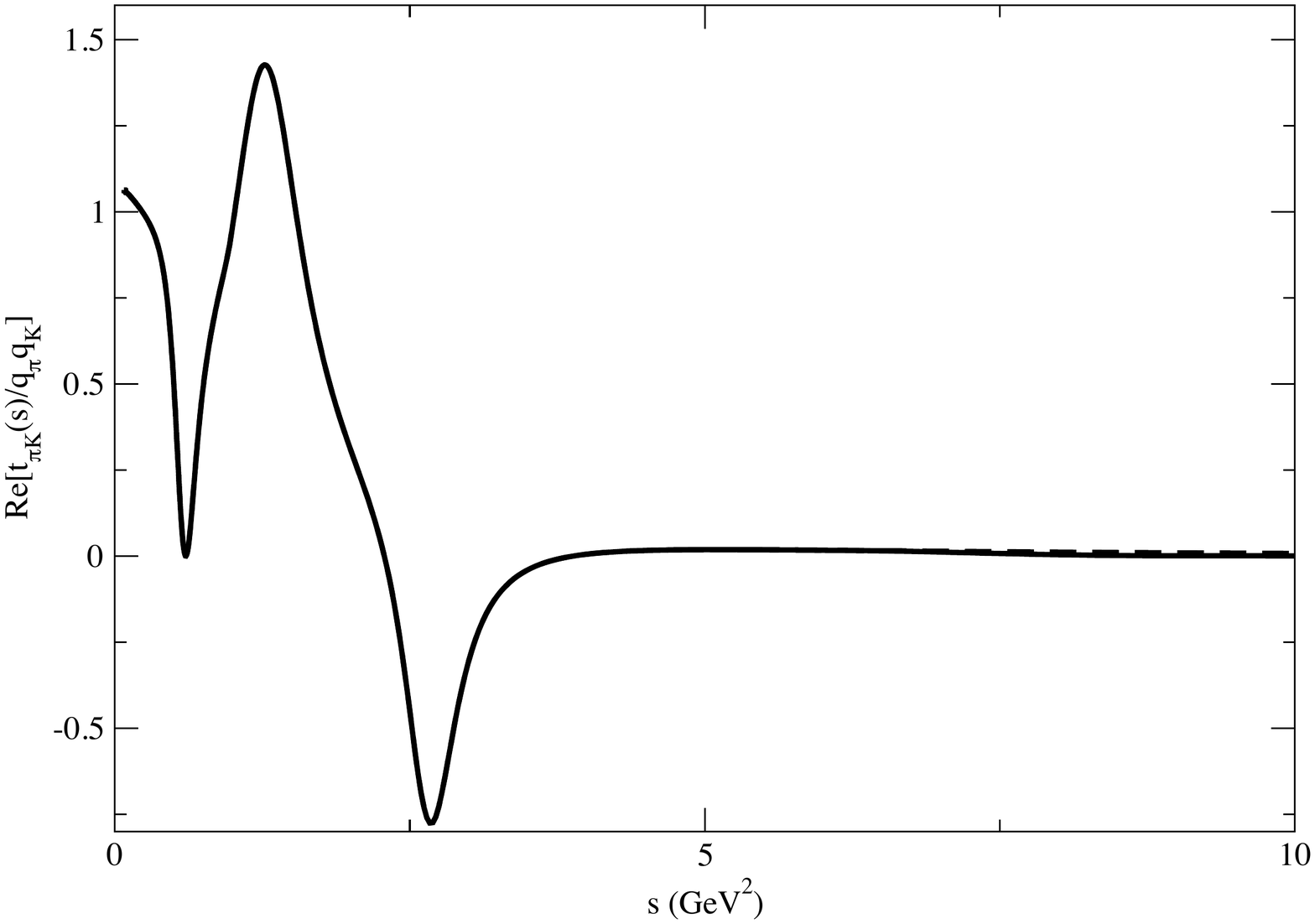} 
\includegraphics[width=3 in,angle=0]{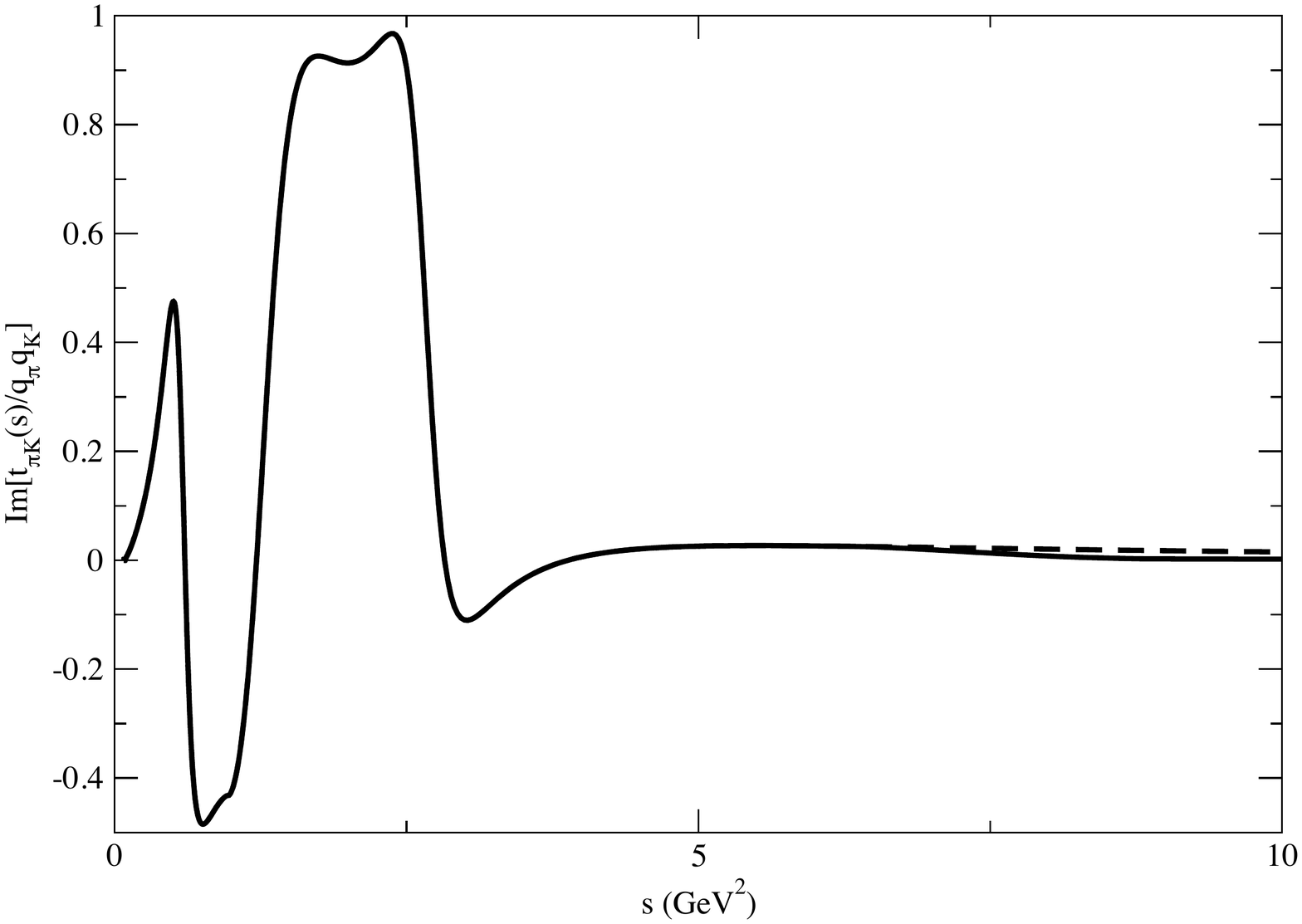}  
\caption{ 
Real (top) and imaginary (bottom) parts of the isovector, $P$-wave amplitude,  $t_{\pi K}(s)/(q_{\pi} q_{K})$ 
 (solid lines). The dashed line is the result of the $K$-matrix parameterization. 
\label{tpikregge}}
\end{center}
\end{figure}

\begin{widetext}
\begin{eqnarray}
t^{Regge}_{\pi K}(s) & \simeq& \frac{1- e^{ - i \pi \alpha_{K^{*}}(0)}}{ \sin  \pi  \alpha_{K^{*}}(0) } \beta_{K^{*}}  \frac{\pi}{8 \sqrt{2}} \frac{2}{3} \int_{-1}^{1} (d \cos \theta )\cos \theta   [   \frac{2 \alpha_{K^{*}}(t)+1}{2 \alpha_{K^{*}}(0)+1} e^{bt} (\alpha'_{K^{*}} s)^{\alpha_{K^{*}}(t)} - (t\rightarrow u)] \nonumber \\
&\simeq&    \frac{1- e^{ - i \pi \alpha_{K^{*}}(0)}}{ \sin  \pi  \alpha_{K^{*}}(0) }   \frac{\pi}{3 \sqrt{2}} \frac{ \beta_{K^{*}}  \alpha'_{K^{*}} }{2 \alpha_{K^{*}}(0)+1} \frac{ (1+2 \alpha_{K^{*}}(0) ) [b+ \alpha'_{K^{*}} \ln(\alpha'_{K^{*}} s)] - 2\alpha'_{K^{*}} }{[b+ \alpha'_{K^{*}} \ln(\alpha'_{K^{*}} s)]^{2}}  (\alpha'_{K^{*}} s)^{\alpha_{K^{*}}(0)-1}  . \nonumber  \\
\end{eqnarray}
\end{widetext}
]The complete amplitude  is given by, 
 \begin{eqnarray}
  &&t_{\pi K}(s) =   \left\{ \begin{array}{c} t_{\pi K}^{Kmatrix}(s)  , s < s_{low}    \\ 
t_{\pi K}^{Regge}(s) ,  s> s_{high}   \end{array}\right.  \ \ \ \  \ \ \ \ 
   \end{eqnarray}
  where we choose $\sqrt{s_{low}}=2.5 $GeV and $\sqrt{s_{high}}=3 $GeV and it is shown in 
   Fig.~\ref{tpikregge}.

\begin{figure}[hh]
\begin{center}
\includegraphics[width=3 in,angle=0]{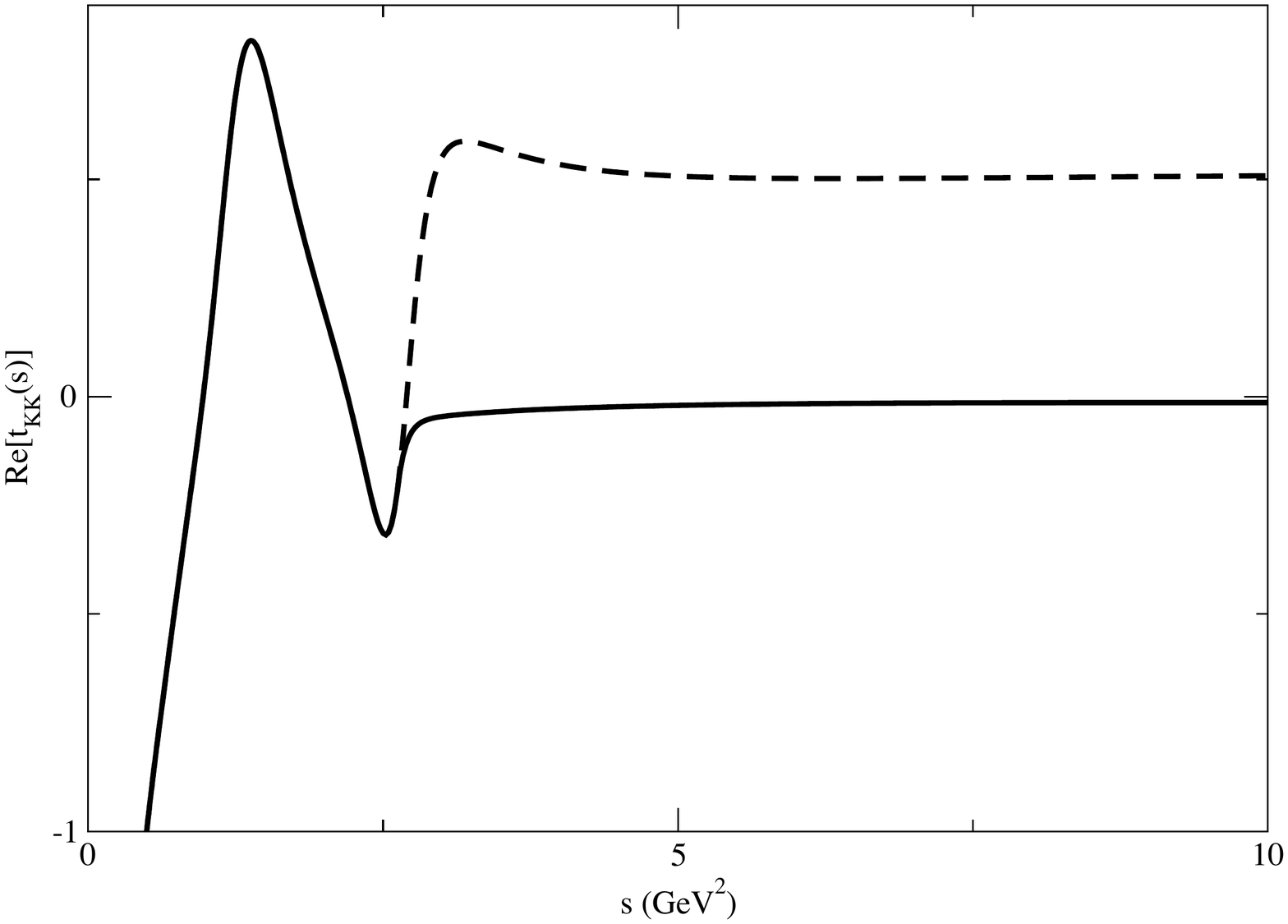} 
\includegraphics[width=3 in,angle=0]{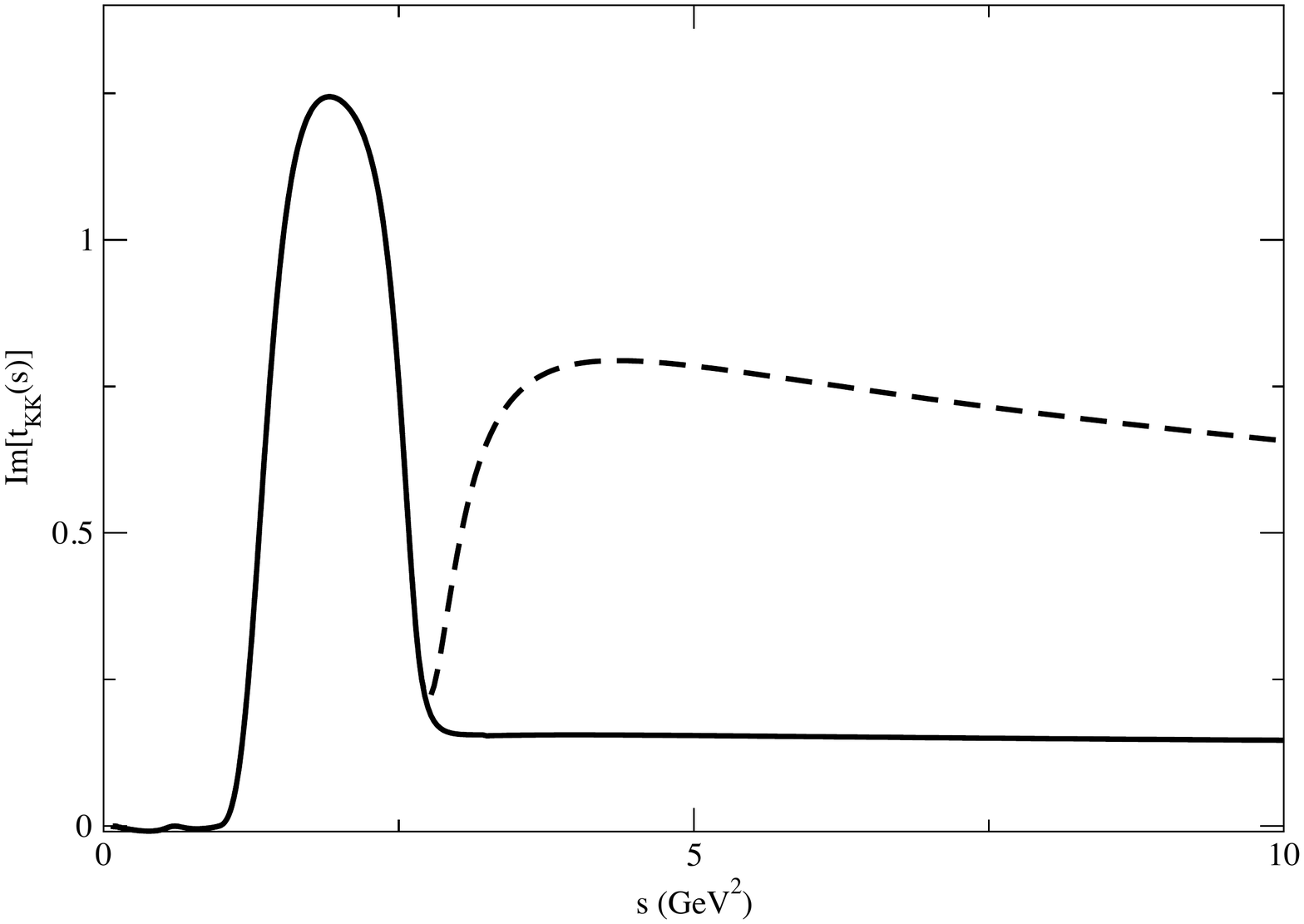}  
\caption{   
Real (top) and imaginary (bottom) parts of the isovector, $P$-wave amplitude,  $t_{K \bar K}(s)$ 
 (solid lines). The dashed line is the result of the $K$-matrix parameterization. 
\label{tkkregge}}
\end{center}
\end{figure}

\begin{itemize} 
\item{} $K \bar K \to K \bar K$ 
\end{itemize} 
  Asymptotically we only retain the Pomeron exchange,  
 \begin{eqnarray}
&& F^{(I_{t}=0)}_{K \bar{K}}(t,s,u)  \nonumber \\
& = & - \frac{1+e^{-i \pi \alpha_{P}(t)}}{\sin \pi \alpha_{P}(t)} \beta^{K\bar{K}}_{P} \alpha_{P}(t) \frac{1+\alpha_{P}(t)}{2} e^{bt}(s/\hat{s})^{\alpha_{P}(t)}  .  \nonumber \\
  \end{eqnarray}
 with  $\alpha_{P}(t)=\alpha_{P}(0)+ t \alpha'_{P}$ and all other parameters, except  $\beta^{K\bar{K}}_{P}$ taken from \cite{pelaez:2005,pelaez:2004}, while for the Pomeron coupling to $K{\bar K}$ we use   relation   $\beta^{K\bar{K}}_{P} = (\frac{ f^{(P)}_{K}}{f^{(P)}_{\pi}})^{2}  (f_{\pi}^{(P)})^{2}= 1.15 $, where the values of $\frac{ f^{(P)}_{K}}{f^{(P)}_{\pi}}  $ and $\beta_{P} =(f^{(P)}_{\pi})^{2}$ are taken from   \cite{pelaez:2005,pelaez:2004}.  
 From $s\leftrightarrow t$ crossing, 
  \begin{eqnarray}
  F^{(I_{s}=1)}_{K \bar{K}}(s,t,u) = \frac{1}{2} F^{(I_{t}=0)}_{K \bar{K}}(t,s,u)    - \frac{1}{2}  F^{(I_{t}=1)}_{K \bar{K}}(t,s,u).  \nonumber \\
  \end{eqnarray}
For the Pomeron contribution to the $s$-channel $P$-wave we thus find 
\begin{eqnarray}
t^{Regge}_{K\bar{K}}(s)  = \frac{\pi}{8}\int_{-1}^{1} ( d \cos \theta  ) \cos \theta     \frac{1}{2} F^{I_{t}=0}_{K \bar{K}}(t,s,u)   .  
\end{eqnarray}
 Asymptotically, $ t^{Regge}_{ K\bar{K}}(s)$ is given by 
  \begin{eqnarray}
&& t^{Regge}_{K \bar{K}}(s)  \nonumber \\
&\simeq & i \frac{\pi}{16} \beta^{K\bar{K}}_{P}   \int_{-1}^{1}(d \cos  \theta)   \cos \theta      \alpha_{P}(t) \frac{1+\alpha_{P}(t)}{2} e^{bt}s^{\alpha_{P}(t)}    \nonumber \\
&\simeq & i   \frac{\pi}{16}\beta^{K\bar{K}}_{P}  \frac{ - 3 \alpha'_{P} +2 (b+\alpha'_{P} \ln s)  }{(b+\alpha'_{P} \ln s)^{2}} s^{\alpha_{P}(0)-1}   . 
 \end{eqnarray}
In   the full  amplitude, 
\begin{eqnarray}
  &&t_{K\bar{K}}(s)  =  \left\{ \begin{array}{c} t_{K\bar{K}}^{Kmatrix}(s)  , s < s_{low}    \\ 
t_{K\bar{K}}^{Regge}(s) ,  s> s_{high}   \end{array}\right. .
   \end{eqnarray}   
  we take  $\sqrt{s_{low} } =1.62 $ GeV and $\sqrt{s_{high} } =3 $ GeV for real parts of amplitudes, and  $\sqrt{s_{low} } =1.64 $ GeV and $\sqrt{s_{high} } =1.8$ GeV for imaginary parts of amplitudes.  The different choice  for the real and imaginary parts allows for a smoother connection with the Regge asymptotics. 
 The phase of  $t_{K\bar{K}}^{Regge}(s) $ asymptotically approaches $\pi/2$ 
  but the phase of $ t_{K\bar{K}}^{Kmatrix}(s)  $  has a sharp drop above  $1.65$ GeV (see right plot in Fig.~\ref{fig:phippregge}). Therefore, choosing  $\sqrt{s_{low} } \sim 1.64 $ GeV allows for a continuous match between the phases of $t_{K\bar{K}}(s)$,  as show in Fig.~\ref{tkkregge}.    

\subsection{Phases of amplitudes and $D$  functions }   \label{ampsregge}
\begin{figure}[hh]
\begin{center}
\includegraphics[width=3 in,angle=00]{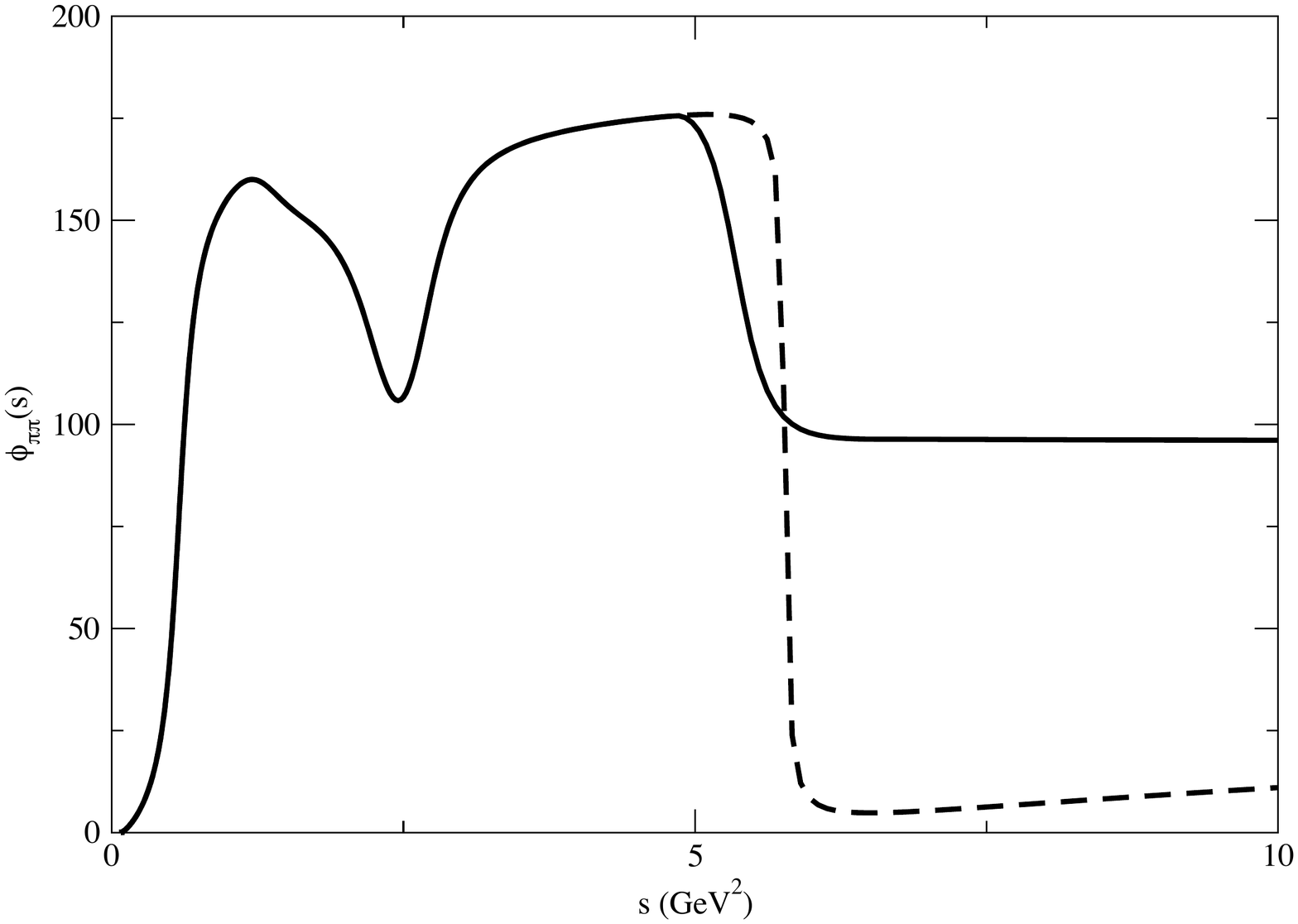}  
\includegraphics[width=3 in,angle=00]{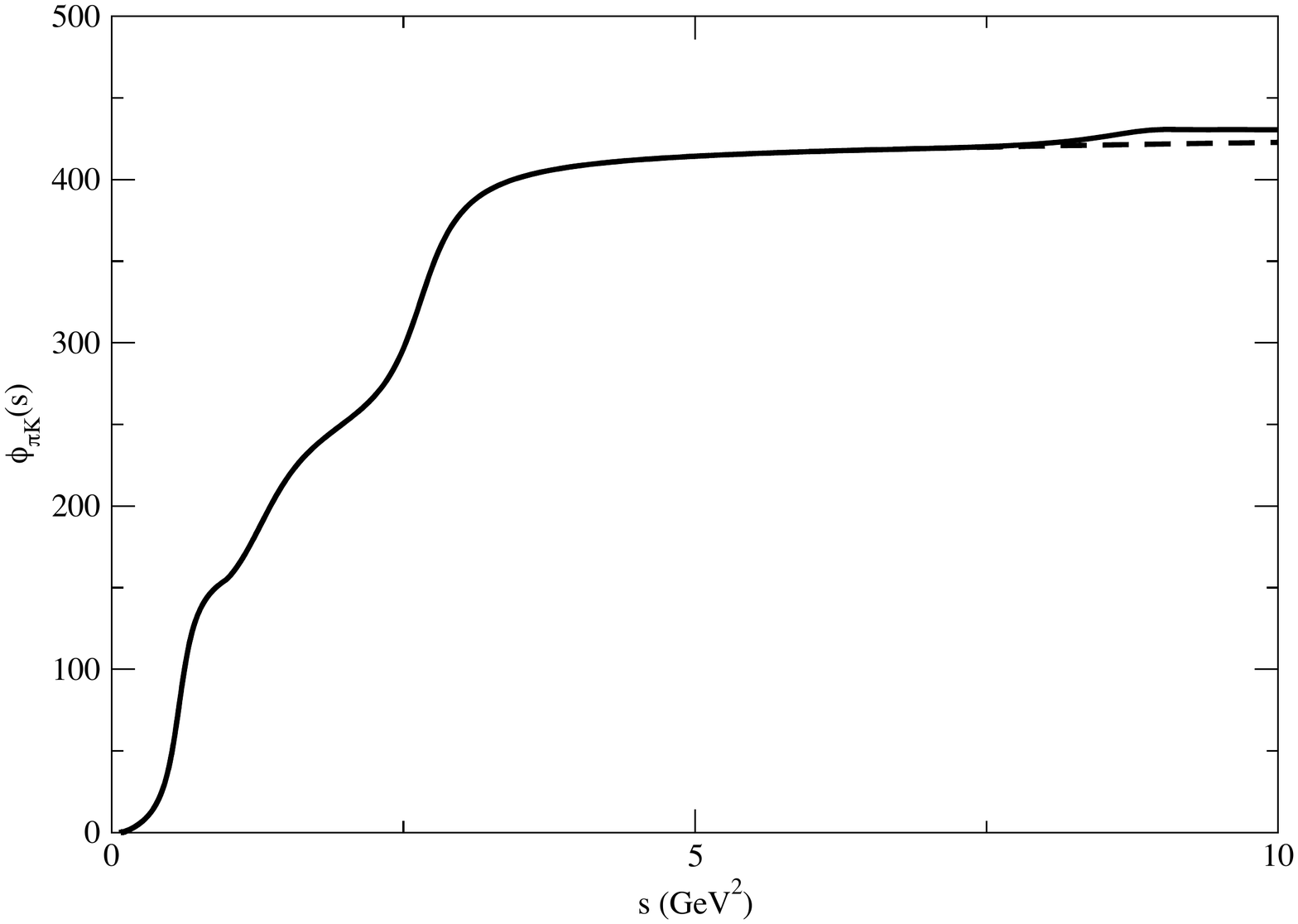} 
\includegraphics[width=3 in,angle=00]{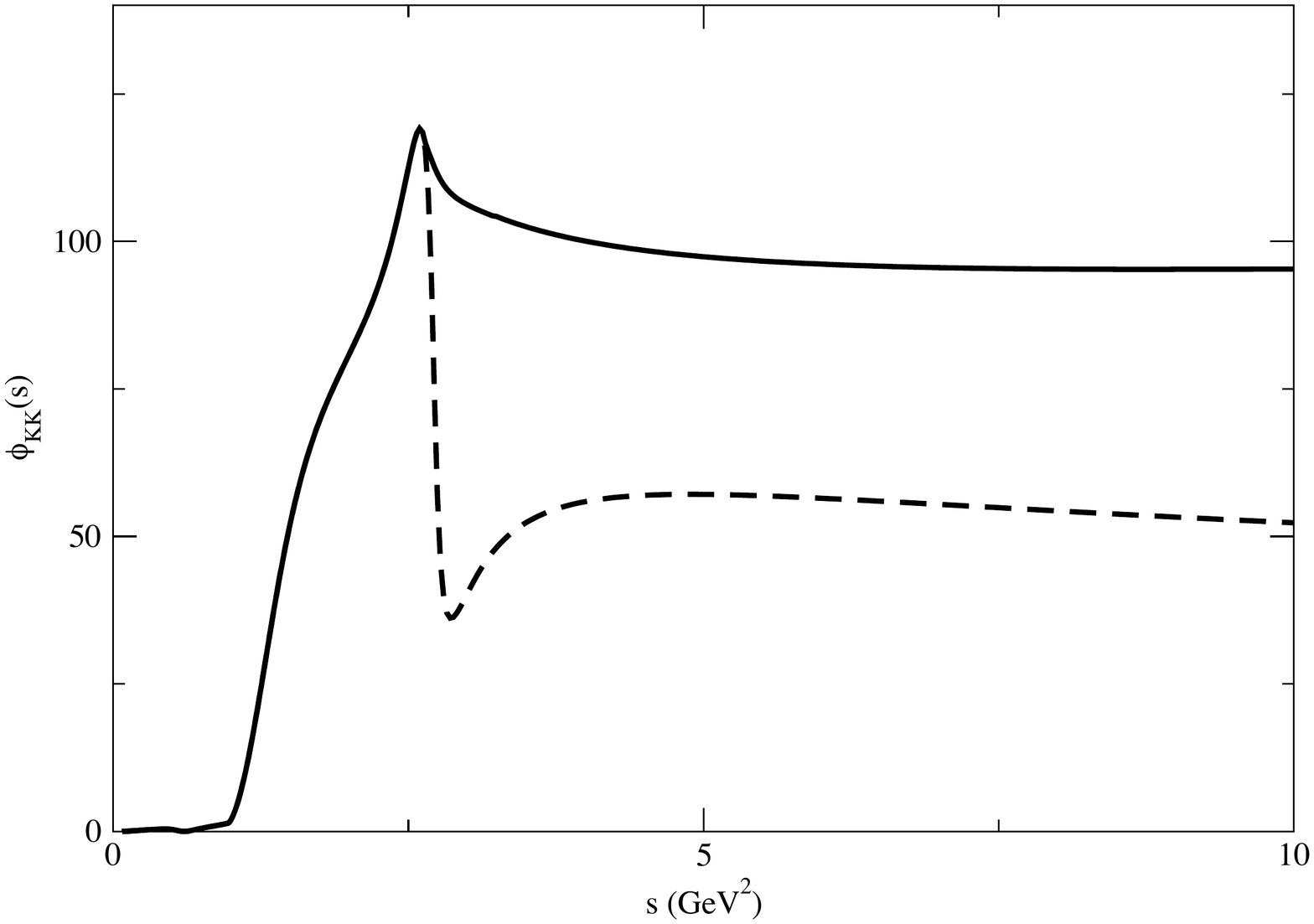}  
\caption{ Phase of the $\pi\pi$ (top), $\pi K$ (middle) and $K \bar{K}$ (bottom) $P$-wave amplitude. 
 The dashed line is the result of the $K$-matrix parameterization from Eq.~(\ref{KND}). 
   \label{fig:phippregge}}
\end{center}
\end{figure}

From  Regge parameterizations we find the following asymptotic behavior for the phases 
 $\phi_{\alpha\beta}(s)$ of the complete amplitudes, 
 \begin{eqnarray}
&& \phi_{\pi \pi} \rightarrow \arctan [-\frac{\sin \pi \alpha_{P}(0)}{1+\cos \pi \alpha_{P}(0)} ]= \frac{\pi}{2}, \\
&&\phi_{\pi K} \rightarrow 2\pi + \arctan [\frac{\sin \pi \alpha_{K^{*}}(0)}{1-\cos \pi \alpha_{K^{*}}(0)} ]  \simeq 2\pi +\frac{\pi}{3}, \ \ \  \  \ \ \ \   \\
&& \phi_{K \bar{K}} \rightarrow  \arctan [-\frac{\sin \pi \alpha_{P}(0)}{1+\cos \pi \alpha_{P}(0)} ] =\frac{\pi}{2}.  
\end{eqnarray}
 These are shown  in Fig.~\ref{fig:phippregge}. 
For the $D$ functions test lead to the  following  asymptotic limits ({\it cf.} Eq.~(\ref{dgen}))

\begin{eqnarray}
&& \frac{ 1  }{D_{\pi \pi}(s)} \rightarrow  \frac{i}{s^{\frac{1}{2}}}, \; 
    \frac{ 1  }{D_{\pi K}(s)} \rightarrow  \frac{1- e^{ - i \pi \alpha_{K^{*}}(0)}}{ \sin  \pi  \alpha_{K^{*}}(0) } \frac{1}{s^{2+\frac{1}{3}}}, \nonumber \\
    && 
   \frac{ 1  }{D_{K \bar{K}}(s)} \rightarrow \frac{i}{s^{\frac{1}{2}}}.
\end{eqnarray}\\

\section{Analytical model for the $P$-wave $\pi K \to \pi K$ amplitude} 
 
 \subsection{$K$-matrix parameterization, ($s < s_{low}$)} \label{kmatofKpiamp}

\begin{figure}[hh]
\begin{center}
\includegraphics[width=3 in,angle=00]{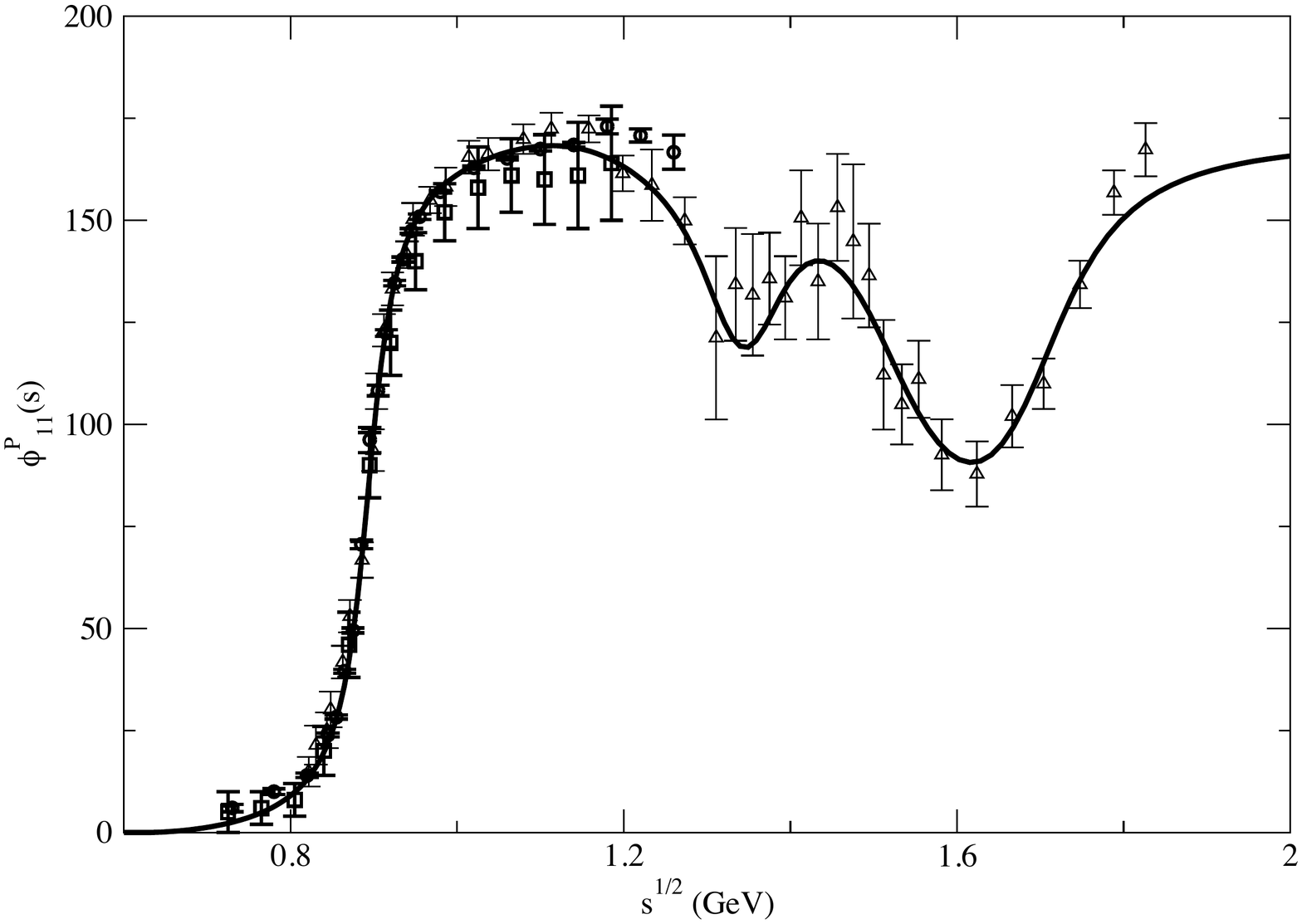}  
\includegraphics[width=3 in,angle=00]{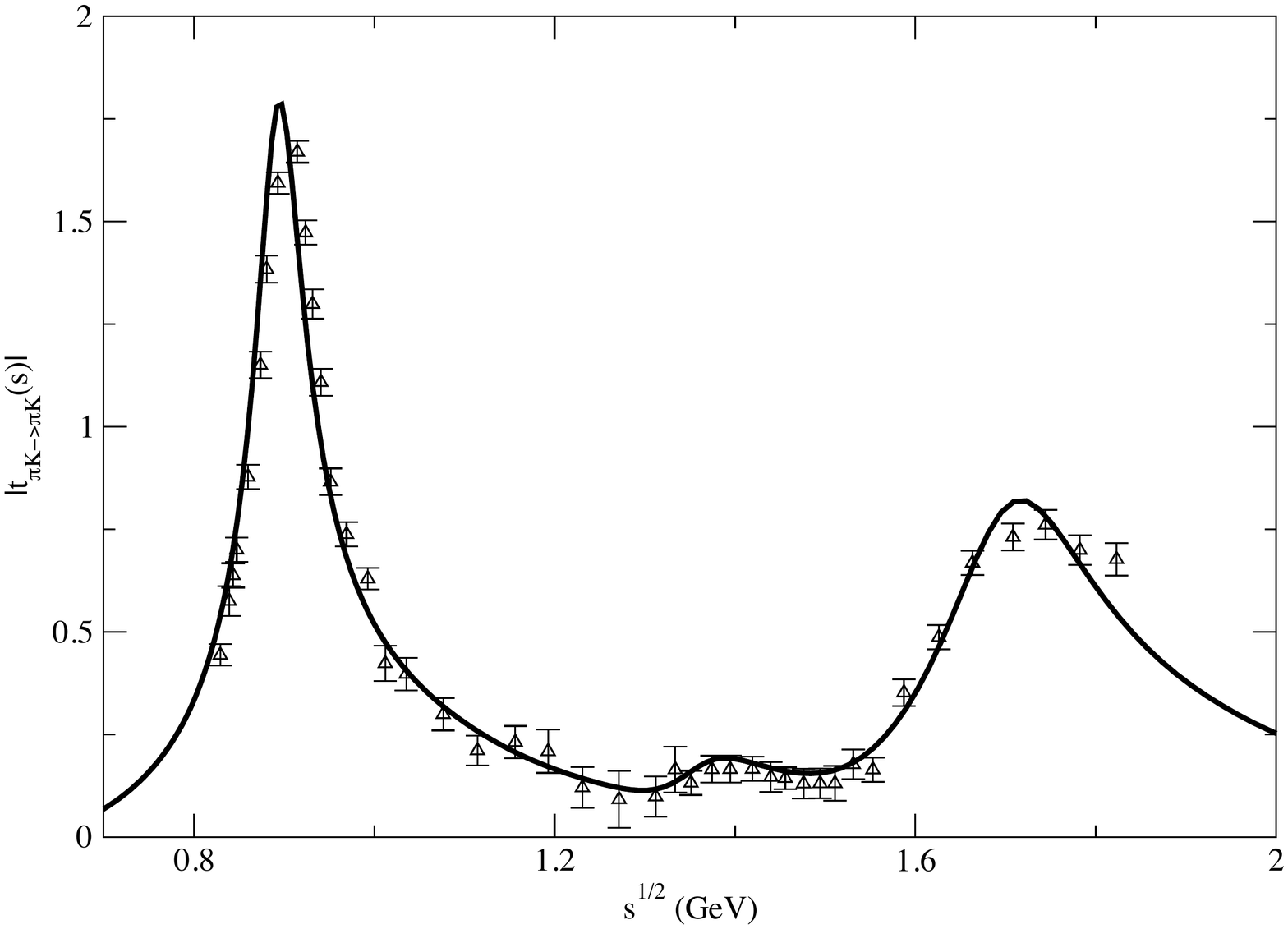}  
\caption{  $\phi_{11}$ (top) and  $|t_{11}|$ (bottom) of $K \pi \rightarrow K\pi$ scattering amplitude vs data from \cite{Mercer:1971} (squares), \cite{Estabrooks:1978} (circles), and  \cite{Aston:1988} (triangles). 
\label{fig:piktopikdata}}
\end{center}
\end{figure}

 To fit the phase shift data on $\pi K$ scattering we use a two-channel $K$-matrix model,
   with the two channels being $K \pi$ and $K^{*}(892) \pi$, and in the second channel 
    treat the $K^*$ as a stable particle, ({\it i.e.} we ignore cuts on the third sheet).
 Similarly to the $\pi\pi$, $K{\bar K}$ case for the $K$-matrix representation of  $K \pi$  and $K^{*}(892) \pi$ amplitudes we write 
 \begin{equation} 
[ \hat t^{-1}(s)]_{\alpha\beta} = [K^{-1}(s)]_{\alpha\beta} + \delta_{\alpha\beta}  \frac{(s - s_\alpha^{+})(s-s_{\alpha}^{-})}{ s}I_\alpha(s),   \label{tkpiktopik}
\end{equation} 
where  $\hat t_{\alpha\beta} \equiv  t_{\alpha\beta}/(4 q_{\alpha} q_{\beta})$,
\begin{eqnarray}
q_{\alpha} & = & \sqrt{ \frac{(s - s_\alpha^{+})(s - s_\alpha^{-})}{4 s}},  s^{\pm}_{\alpha} = (m_{\alpha} \pm m_{\pi})^{2}, \nonumber \\
& &  m_{1}=m_{K}, m_{2} = M_{K^{*}(892)}, 
\end{eqnarray}
and
\begin{equation}  \label{kpiloop}
I_\alpha(s) = I_\alpha(0) - \frac{s}{\pi} \int_{s_\alpha^{+}}^\infty ds' \frac{  \sqrt{(1 - \frac{s_\alpha^{+}}{s'})(1 - \frac{s_\alpha^{-}}{s'})} }{ s'(s' - s)}.  
\end{equation} 
 A convenient choice for the subtraction constant, $I_\alpha(0)$, is to take $\mbox{Re} I_\alpha(M^{2}_{K^{*}(892)}) = 0$ so that one of the poles of  $K_{11}$  is located at  mass squared of the $K^*(892)$, $m_2^2$.  In terms of phase shift and inelasticity the $K\pi$ and $K^{*}(892) \pi$ amplitudes are given by 
  \begin{eqnarray} 
& &  t_{11} = \frac{\eta e^{2i\delta_{11}} - 1}{2i \rho_{1} },  t_{22 } = \frac{\eta e^{2i\delta_{22}} - 1}{2i \rho_{2}},   \nonumber \\
 & & t_{12} = t_{21}  = \frac{\sqrt{1 - \eta^2} e^{i (\delta_{11} + \delta_{22})} }{2\sqrt{\rho_{1} \rho_{2}}} 
  \end{eqnarray} 
where $\rho_{\alpha}(s) =\sqrt{(1-\frac{s_{\alpha}^{+}}{s})(1-\frac{s_{\alpha}^{-}}{s})}$. 
  The denominator $D_{\alpha\beta}$ of the  $K\pi$ and $K^{*}(892) \pi$
    amplitudes are defined by Omn\'es-Muskhelishvili function
   \begin{equation} \label{kpiampDfunc}
D_{\alpha\beta} (s) = \exp\left( - \frac{s }{\pi} \int_{(m_{K}+m_{\pi})^{2}}^\infty ds' 
 \frac{\phi_{\alpha\beta}(s')}{s'(s' - s) } \right). 
 \end{equation}
    To fit the $P$-wave phase  shift data  \cite{Mercer:1971,Estabrooks:1978,Aston:1988}
     we use a three-pole parameterization of the K-matrix
    \begin{eqnarray}
& & K_{11} =  \frac{ \alpha_1^2 }{M^2_{K^{*}(892)}-s}   +\frac{\beta_{1}^{2} }{s_{2}-s}+\frac{\lambda_{1}^{2} }{s_{3}-s}  + \gamma^{(0)}_{11} +   \gamma^{(1)}_{11}  s, \nonumber \\ 
&& K_{22} =  \frac{ \beta^{2}_{2}}{s_{2}-s} +\frac{\lambda_{2}^{2}  }{s_{3}-s}+ \gamma^{(0)}_{22}+ \gamma^{(1)}_{22} s, \nonumber \\ 
& & K_{12}  = K_{21} =  \frac{\beta_{1} \beta_{2}}{s_{2}-s} +\frac{\lambda_{1} \lambda_{2} }{s_{3}-s} + \gamma^{(0)}_{12}+ \gamma^{(1)}_{12}s , 
 \end{eqnarray}
 where 
 \begin{equation} 
 \alpha_1^2 = \frac{\Gamma_{K^{*}(892)}M^5_{K^{*}(892)}}{[(M_{K^{*}(892)}^2 - s_1^{+})(M_{K^{*}(892)}^2 - s_1^{-})]^{3/2}}.
 \end{equation} 
And for the parameters of the $K$-matrix obtain
  $\Gamma_{K^{*}(892)} = 0.0504 \mbox{ GeV}$, 
\begin{eqnarray} \label{kmatparameters}
&&   \sqrt{ s_{2}}  = 1.35 \mbox{ GeV},  \ \ \ \  \sqrt{ s_{3}}  = 1.75 \mbox{ GeV}, \ \ \ \  \beta_{1} = 0.110,  \nonumber \\
&&   \beta_{2} =-0.685, \ \ \ \ \lambda_{1} =0.142, \ \ \ \ \lambda_{2} =1.089. 
 \nonumber  \\  && \gamma^{(0)}_{11} = 0.204, \ \ \ \ \ \gamma^{(0)}_{12} =-0.983, \ \ \ \ \ \gamma^{(0)}_{22} = 8.329,  \nonumber \\
 && \gamma^{(1)}_{11} = -0.052, \ \ \ \ \ \gamma^{(1)}_{12} =  0.426, \ \ \ \ \ \gamma^{(1)}_{22} = -3.834.   \nonumber \\
 \end{eqnarray}
  with the $\gamma^{(0)}$'s in units of $\mbox{ GeV}^{-2}$ and  $\gamma^{(1)}$'s in units of $\mbox{ GeV}^{-4}$. The phase, $\phi_{11}$ and  magnitude,  $|t_{11}|$  of $K \pi \rightarrow K\pi$ scattering amplitude is compared to the date in Fig.~\ref{fig:piktopikdata}. 
      We can express $\hat{t}_{\alpha\beta}= t_{\alpha\beta}/(4 q_{\alpha} q_{\beta})$  in terms of a product of poles, zeros and the Omn\'es-Muskhelishvili function
        \begin{eqnarray} 
& &  \hat{t}_{\alpha\beta} =N_{\alpha\beta} \frac{\prod_{l=1,N_{z,\alpha\beta}} (s-s^{(\alpha\beta)}_{z,l})}{\prod_{l=1,8} (s-s_{P,l})} \nonumber \\
&\times& \exp\left(  \frac{s }{\pi} \int_{(m_{K}+m_{\pi})^{2}}^\infty ds' 
 \frac{\phi_{\alpha\beta}(s')}{s'(s' - s) } \right). 
  \end{eqnarray} 
 where $N_{z,\alpha\beta}$ is the number of zeros of $\hat{t}_{\alpha\beta}$ for which we find 
  $N_{z,11}=N_{z,22}=7$, $N_{z,12}=N_{z,21}=6$. 
 The  normalization factors are given by  $N_{11}=7.075, N_{12} =-421.989, N_{22}=2.808$. 
  The positions of the poles and zeros are given by (in units of $\mbox{GeV}^2$)  
      \begin{eqnarray}  \label{kmatpoles}
& & s_{P,1/2} =0.3573 \pm i 0.4055, \ \ s_{P,3/4} =2.4912 \pm i 0.5762,  \nonumber \\
&&  s_{P,5/6} =20.3504 \pm i 3.3856,  \nonumber \\
&& s_{P,7} =-0.00489, \ \  s_{P,8} =-6.2666, 
  \end{eqnarray} 
and 
     \begin{eqnarray} \label{kmatzeros}
& & s^{(11)}_{z,1/2} =0.3428 \pm i 0.4470, \ \  s^{(11)}_{z,3/4}=2.2188 \pm i 0.6175,  \nonumber \\
&&  s^{(11)}_{z,5/6}=12.8061 \pm i 0.2470,  \ \ s^{(11)}_{z,7} =0. \nonumber \\
&& s^{(12)}_{z,1/2} =1.9181 \pm i 0.3669, \ \  s^{(12)}_{z,3}=3.3956, \nonumber \\
&& s^{(12)}_{z,4}=M_{K^{*}(892)}^{2},  \ \ s^{(11)}_{z,5/6} =0. \nonumber  \\
& & s^{(22)}_{z,1/2} =2.2704 \pm i 0.2273, \ \  s^{(22)}_{z,3/4}=20.2775 \pm i 3.3479, \nonumber \\
&&  s^{(22)}_{z,5}=-0.00489,  \ \  s^{(22)}_{z,6}=-6.1874,  \ \ s^{(22)}_{z,7}= 0,  \nonumber \\
  \end{eqnarray} 
  respectively.  
  As can be seen from Fig.~\ref{tpiktopikregge},  this $K$-matrix leads to a dramatic, 
    most likely unphysical,  drop in  the phase $\phi_{11}$ (dashed line)   around  $6 \mbox{ GeV}^{2}$  and this results in  both $\phi_{11}$ and  $\delta_{11}$ vanishing  asymptotically.  
  Furthermore the resulting t-matrix has complex poles and zeros on the physical sheet (see Eq.~(\ref{kmatpoles}) and Eq.~(\ref{kmatzeros})).     The  origin of these unphysical poles 
   can be illustrated by considering a single channel, $K_{11}$ only, with a single pole and constant background term. The resulting $t$-matrix element  is then given by, 
  \begin{widetext}
   \begin{eqnarray}
\hat{t}_{11}&=& \frac{ 1}{8} \frac{ \alpha_{1}^{2} + \gamma^{0}_{11} (M_{K^{*}(892)}^{2}-s) }{(M_{K^{*}(892)}^{2}-s) -  \frac{[s-(m_{K}-m_{\pi})^{2}][s-(m_{K}+m_{\pi})^{2}]}{ s }I_{1}(s) [ \alpha_{1}^{2} + \gamma^{0}_{11} (M_{K^{*}(892)}^{2}-s) ] }  .  
 \end{eqnarray}
 \end{widetext}
If,  for simplicity, we  replace $m_{K}$ by $m_{\pi}$ and  keep only 
  the imaginary of $I_{1}(s)$, in the limit $|s| \rightarrow \infty$ and $\gamma^{0}_{11} \rightarrow 0$  with $|\gamma^{0}_{11}| |s| \gg \alpha_{1}^{2} $ one finds 
    \begin{eqnarray}
\hat{t}_{11}&\rightarrow& \frac{ 1}{8} \frac{   \gamma^{0}_{11}  }{1 -  i  \sqrt{1- \frac{ 4 m_{\pi}^{2}}{s}} \gamma^{0}_{11} s  }  .  
 \end{eqnarray}
In  the limit  $\gamma^{0}_{11} \rightarrow 0$  the pole is on the first sheet  and approaches 
  $s\rightarrow  \pm \frac{i}{\gamma^{0}_{11} } $. 
   Even though the K-matrix itself has unphysical singularities and zeros it still faithfully reproduces the phase and magnitude of the amplitude of $K\pi$ scattering data up to $1.8\mbox{ GeV}$. Similarly to the cases in $\pi \pi $ and $K\bar{K}$ scattering presented in previous sections, we will truncate the K-matrix solution at $s_{low}$ and match it with  Regge parameterization at $s_{high}$.

\subsection{Regge parameterization  for $\pi^{0} K^{\pm } \rightarrow \pi^{0} K^{\pm}$ ($s>s_{high}$)} \label{reggepiktopik}

\begin{figure}[hh]
\begin{center}
\includegraphics[width=3 in,angle=0]{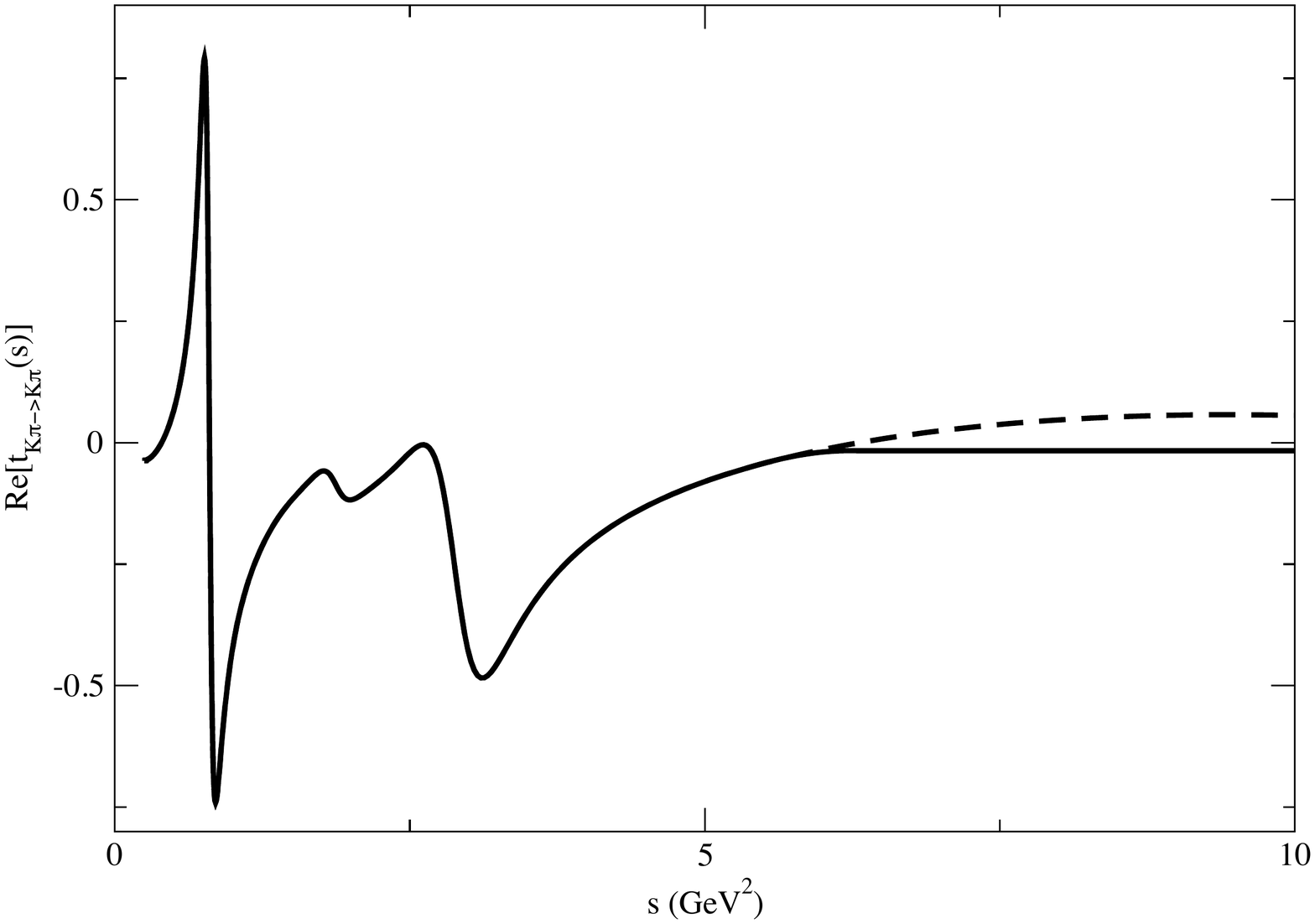} 
\includegraphics[width=3 in,angle=0]{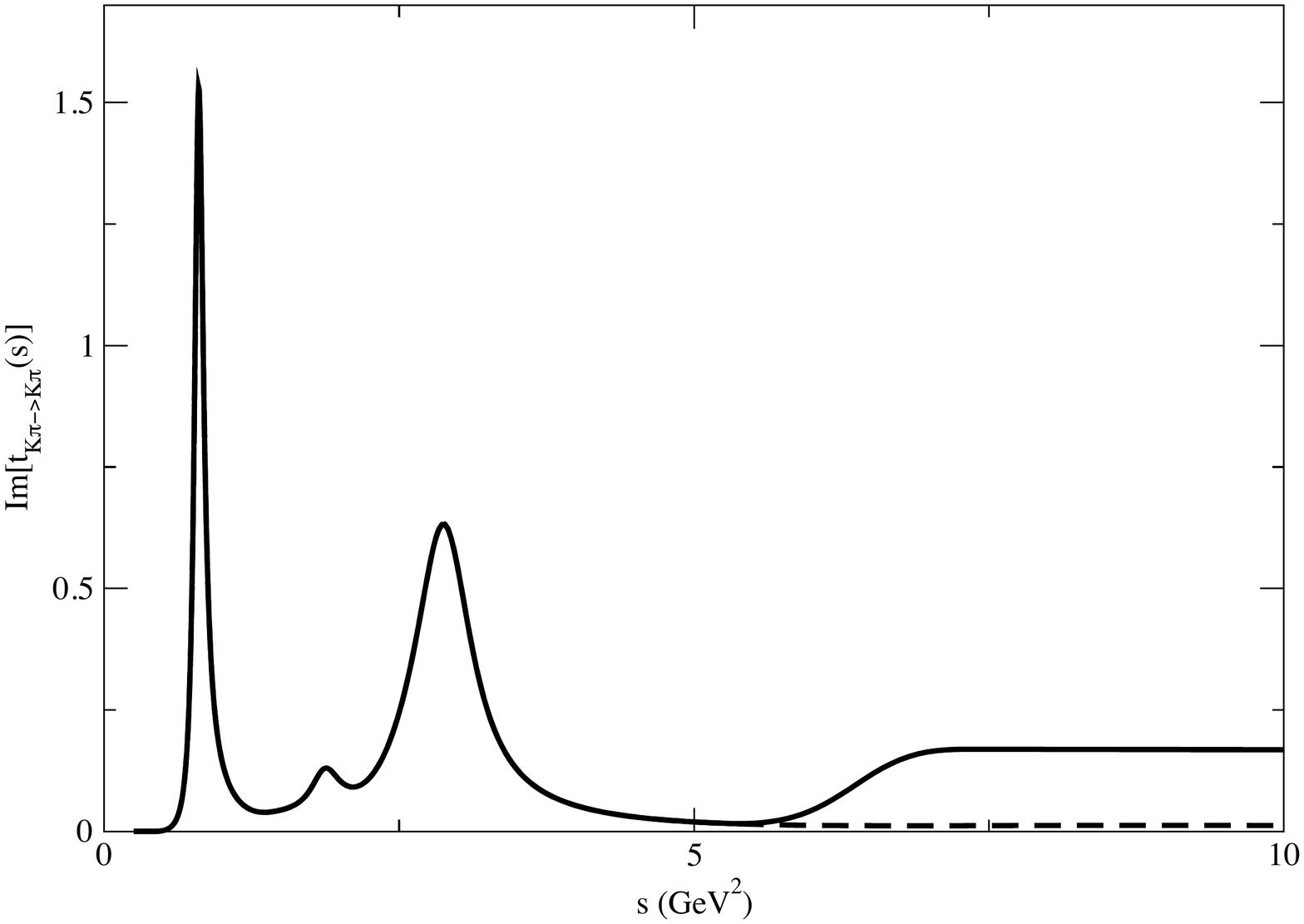}  
\includegraphics[width=3 in,angle=00]{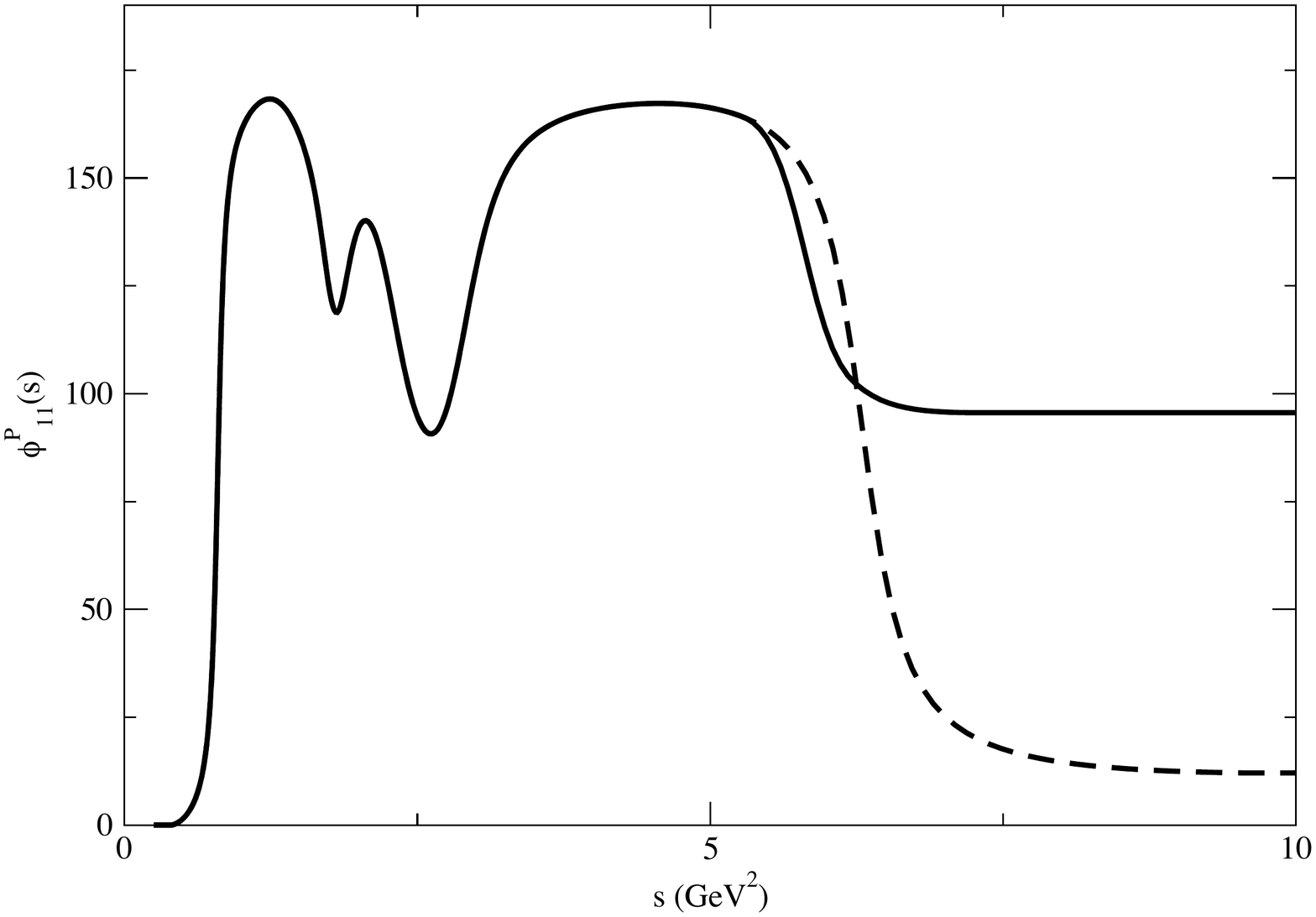}  
\caption{   
Real (left) and imaginary (right) parts of the isovector, $P$-wave amplitude,  $t_{\pi K \rightarrow \pi K}(s)$  and phase  $\phi_{11}$
 (solid curves). The dashed curves are the result of the $K$-matrix parameterization. 
\label{tpiktopikregge}}
\end{center}
\end{figure}

Asymptotically  we only retain Pomeron ($P$) in the t-channel and the  $K^{*}$ trajectory in the u-channel
\begin{eqnarray}
&& F^{I_{t}=0}_{\pi  K \rightarrow  \pi K   }(t,s, u) \nonumber \\
 &=& - \frac{1+e^{-i \pi \alpha_{P}(t)}}{\sin \pi \alpha_{P}(t)}   \beta^{\pi K}_{P} \alpha_{P}(t) \frac{1+\alpha_{P}(t)}{2} e^{bt}(s/\hat{s})^{\alpha_{P}(t)}. \nonumber \\
 &&  \\
&& F^{I_{u}= \frac{1}{2}}_{\pi  K \rightarrow \pi K} (u,t,s)  \nonumber \\
 &=&   \frac{1- e^{ - i \pi \alpha_{K^{*}}(u)}}{ \sin  \pi  \alpha_{K^{*}}(u) } \beta_{K^{*}} \frac{2 \alpha_{K^{*}}(u)+1}{2 \alpha_{K^{*}}(0)+1} e^{bu} (\alpha'_{K^{*}} s)^{\alpha_{K^{*}}(u)}. \nonumber \\
 \end{eqnarray}
The Pomeron trajectory is  given in Eq.~(\ref{pomerontraject}) with  parameters in Pomeron parameterization are  taken from \cite{pelaez:2005,pelaez:2004}, except the coupling constant  
     $ \beta^{\pi K}_{P} = f^{(P)}_{\pi} f^{(P)}_{K} =[ \beta_{P}^{\pi \pi} \beta_{P}^{K\bar{K}}]^{\frac{1}{2}}=1.709$. The $K^{*}$ trajectory is given by  $\alpha_{K^{*}}(u) =0.352+0.882 u$ as in Section \ref{appendregge}.   From s-t and s-u channel crossing, we obtain
  \begin{eqnarray}
&& F^{I_{s}=\frac{1}{2}}_{\pi  K \rightarrow \pi K }(s,t,u)  \nonumber \\
&=& \frac{1}{\sqrt{6}}F^{I_{t}=0}_{\pi  K \rightarrow \pi K}(t,s, u)+F^{I_{t}=1}_{\pi  K \rightarrow \pi K }(t,s, u) \nonumber \\
& +&  \frac{1}{3}F^{I_{u}= \frac{1}{2}}_{\pi  K \rightarrow \pi K} (u,t,s) +  \frac{4}{3}F^{I_{u}= \frac{3}{2}}_{\pi  K \rightarrow \pi K } (u,t,s). \nonumber \\
\end{eqnarray}
The $P$-wave projection of the Regge amplitude in $\pi K \rightarrow \pi K$ scattering is given by
\begin{eqnarray}
&& t^{Regge}_{\pi K \rightarrow \pi K} (s) =  \frac{\pi}{8}\int_{-1}^{1} ( d \cos) \cos \theta  \nonumber \\
&\times&  [ \frac{1}{\sqrt{6}}F^{I_{t}=0}_{\pi  K \rightarrow \pi K}(t,s,u)+ \frac{1}{3}F^{I_{u}= \frac{1}{2}}_{\pi K \rightarrow  \pi K } (u,t,s)]    .   \ \ \ \  \ \ \ \ 
\end{eqnarray}\\
The complete amplitude for the  $\pi K \rightarrow \pi K$ amplitude is given by, 
 \begin{eqnarray}
  &&t_{ \pi K \rightarrow \pi K}(s) =   \left\{ \begin{array}{c} t_{11}^{Kmatrix}(s)  , s < s_{low}    \\ 
t_{\pi K \rightarrow \pi K}^{Regge}(s) ,  s> s_{high}   \end{array}\right.  \ \ \ \  \ \ \ \ 
   \end{eqnarray}
  where we choose $\sqrt{s_{low}}=2.3 $GeV,  $\sqrt{s_{high}}=2.5 $GeV for the real part of the amplitude and  $\sqrt{s_{high}}=2.7 $GeV for the imaginary part of the amplitude as shown in 
   Fig.~\ref{tpiktopikregge}. 
From  the $P$-wave projection of the  Regge amplitude we find the following asymptotic behavior for the phase  $\phi_{11}(s)$ and  denominator function of the complete amplitudes, 
 \begin{eqnarray}
&& \phi_{11} \rightarrow \arctan [-\frac{\sin \pi \alpha_{P}(0)}{1+\cos \pi \alpha_{P}(0)} ]= \frac{\pi}{2},  \nonumber \\
&& \frac{ 1  }{D_{\pi K \rightarrow \pi K}(s)} \rightarrow  \frac{i}{s^{\frac{1}{2}}} 
\end{eqnarray}
The phase is shown  in Fig.~\ref{tpiktopikregge},


\begin{thebibliography}{99}






 
 














  \bibitem{Diekmann:1988}
B. Diekmann,
  Phys.\ Rep.\  {\bf 159}, 99 (1988).




     \bibitem{Hyams:1973}
B. Hyams, C. Jones and P. Weilhammer,
  Nucl.\ Phys.\  B {\bf 64}, 134 (1973).


  \bibitem{Aston:1980}
D. Aston {\it et al.},
  Phys.\ Lett.\  B {\bf 92}, 215 (1980).

  \bibitem{Bisello:1989}
D. Bisello {\it et al.},
  Phys.\ Lett.\  B {\bf 220}, 321 (1989).

  \bibitem{Donnachie:1987}
A. Donnachie and H. Mirzaie,
  Z.\ Phys.\  C {\bf 33}, 407 (1987).





  \bibitem{Jozef:2010}
J. J. Dudek, R. G. Edwards, M. J. Peardon, D. G. Richards, and C. E. Thomas, 
  Phys.\ Rev.\  D {\bf 82}, 034508 (2010).



  \bibitem{Isgur:1985}
S. Godfrey and N. Isgur ,
  Phys.\ Rev.\  D {\bf 32}, 189 (1985).




  \bibitem{PDG}
K. Nakamura   {\it et al.}  (Particle Data Group),
  J.\ Phys.\  G {\bf 37}, 075021 (2010).
 
 
 
 \bibitem{Achasov:1997}
N.N.Achasov and A.A.Kozhevnikov  ,
 Phys.\ Atom.\  Nucl. {\bf 60}, 1011 (1997). 




 \bibitem{Achasov:2002}
N.N.Achasov and A.A.Kozhevnikov  ,
 Phys.\ Atom.\  Nucl. {\bf 65}, 153 (2002). 



 
 
    
\bibitem{Guo:2010}
  P.~Guo, R.~Mitchell and A.~P.~Szczepaniak,
  Phys.\ Rev.\  D {\bf 82}, 094002 (2010).

 
  
  \bibitem{BES:2006}
BES Collaboration (M. Ablikim {\it et al.}),
  Phys.\ Rev.\  Lett. {\bf 97}, 142002 (2006).



  \bibitem{Liu:2007}
X. Liu,  B. Zhang, L.-L. Shen, and S.-L. Zhu, 
  Phys.\ Rev.\  D {\bf 75}, 074017 (2007).



  \bibitem{Li:2007}
B. A. Li, 
  Phys.\ Rev.\  D {\bf 76},  094016 (2007).



\bibitem{Szczepaniak:2010re}
  A.~P.~Szczepaniak, P.~Guo, M.~Battaglieri and R.~De Vita,
  Phys.\ Rev.\  D {\bf 82}, 036006 (2010)
  
        


 

  
   \bibitem{Protopopescu:1973}
S.D. Protopopescu {\it et al.} ,
  Phys.\ Rev.\  D {\bf 7}, 1279 (1973).

       \bibitem{Estabrooks:1974}
P. Estabrooks and A. D. Martin,
  Nucl.\ Phys.\  B {\bf 79}, 301 (1974).
  


\bibitem{Pham:1976yi}
  T.~N.~Pham and T.~N.~Truong,
  Phys.\ Rev.\  D {\bf 16}, 896 (1977).
 

   
  \bibitem{Aston:1988}
 D. Aston  {\it et al.},
  Nucl.\ Phys.\  B {\bf 296}, 493 (1988).


  
  
  \bibitem{Mercer:1971}
 R. Mercer  {\it et al.},
  Nucl.\ Phys.\  B {\bf 32}, 381 (1971).
  
  
  
  \bibitem{Estabrooks:1978}
 P. Estabrooks  {\it et al.},
  Nucl.\ Phys.\  B {\bf 133}, 490 (1978).

  
  
    






    
  
  
    

  




     
   

  

   
   
   
   
  
  


   




    
  
    
  
  
  
   \bibitem{pelaez:2004}
 J. R. Pel‡ez   and F. J. Yndur‡in  ,
  Phys.\ Rev.\  D {\bf 69}, 114001 (2004).



  \bibitem{Ananthanarayan:2001}
B. Ananthanarayan, G. Colangelo, J. Gasser   and H. Leutwyler ,
  Phys.\ Rep. {\bf 353}, 207 (2001).


  
  
  
  \bibitem{pelaez:2005}
 J. R. Pel‡ez   and F. J. Yndur‡in  ,
  Phys.\ Rev.\  D {\bf 71}, 074016 (2005).

  
  
  \bibitem{Ananthanarayan:2002}
 B. Ananthanarayan,  P. BŸttiker and  B. Moussallam  ,
  Eur.\ Phys.\ J.\  C {\bf 22}, 133 (2001).



  \bibitem{cohen:1980}
 D. H. Cohen, D.S. Ayres, R. Diebold, S.L. Kramer, A.J. Pawlicki and A.B. Wicklund  ,
  Phys.\ Rev.\  D {\bf 22}, 2595 (1980).

  

  
  
  
  
       
         
      
            
   
    
\end{thebibliography}
\end{document}